\documentclass[useAMS,usenatbib]{mn2e}
\newcommand{\bib}{\bibitem[\protect\citeauthoryear}
\usepackage[dvips]{graphicx}
\usepackage[T1]{fontenc}
\usepackage{ae,aecompl}

\title[B2\,0326+39 \& B2\,1553+24 jet models]{Relativistic models of
two low-luminosity radio jets: \\ B2\,0326+39 and B2\,1553+24}
\author[J.R. Canvin \& R.A. Laing]{J.R. Canvin \thanks{E-mail:
jrc@astro.ox.ac.uk}$^{1}$ , R.A. Laing$^{2,1}$\\ $^1$ University of
Oxford, Department of Astrophysics, Denys Wilkinson Building, Keble
Road, Oxford OX1 3RH \\ $^2$ European Southern Observatory,
Karl-Schwarzschild-Stra\ss e 2, D-85748 Garching-bei-M\"{u}nchen,
Germany, \\ }

\date{Received}

\pagerange{\pageref{firstpage}--\pageref{lastpage}} \pubyear{2003}

\begin{document}
\label{firstpage}

\maketitle

%-------------------------- ****** ABSTRACT ****** --------------------------%
\begin{abstract}
We apply the intrinsically symmetrical, decelerating relativistic jet
model developed by Laing \& Bridle for 3C\,31 to deep, full-synthesis 8.4-GHz
VLA imaging of the two low-luminosity radio galaxies B2\,0326+39 and
B2\,1553+24.  After some modifications to the functional forms used to
describe the geometry, velocity, emissivity and magnetic-field structure,
these models can accurately fit our data in both total intensity and
linear polarization.  We conclude that the jets in B2\,0326+39 and
B2\,1553+24 are at angles of $64^\circ\pm5^\circ$ and
$7.7^\circ\pm1.3^\circ$ to the line of sight, respectively.  In both
objects, we find that the jets decelerate from 0.7 -- 0.8$c$ to $<$0.2$c$
over a distance of approximately 10\,kpc, although in B2\,1553+24 this
transition occurs much further from the nucleus than in B2\,0326+39 or
3C\,31. The longitudinal emissivity profiles can be divided into sections,
each fit accurately by a power law; the indices of these power laws
decrease with distance from the nucleus. B2\,0326+39 also requires a
discontinuity in emissivity to in order to fit a region with several
bright knots of emission.  In B2\,1553+24, the sudden brightening of the
jet can be explained by a combination of rapid expansion of the jet and a
continuous variation of emissivity. The magnetic fields in both objects
are dominated by the longitudinal component in the high-velocity regions
close to the nucleus and by the toroidal component further out, but
B2\,0326+39 also has a significant radial component at large distances,
whereas B2\,1553+24 does not. Simple adiabatic models fail to fit the
emissivity variations in the regions of high velocity but provide good
descriptions of the emissivity after the jets have decelerated. Given the
small angle to the line of sight inferred for B2\,1553+24, there should be
a significant population of similar sources at less extreme
orientations. Such objects should have long ($\ga$200\,kpc), straight,
faint jets and we show that their true sizes are likely to have been
underestimated in existing images.

\end{abstract}


%------------------------ ****** INTRODUCTION ****** ------------------------%
\section{Introduction}
\label{intro}

Evidence that the jets in low-luminosity, FR\,I \citep{FR74} radio
galaxies are initially relativistic and decelerate on kpc scales has
mounted in recent years. FR\,I sources are thought to be the side-on
counterparts of BL Lac objects, in which relativistic motion on parsec
scales is well-established \citep{UP95}.  Superluminal motions have been
seen on milliarcsecond scales in several FR\,I jets \citep{Giovannini01}
and on arcsecond scales in M\,87 \citep{Biretta95}. In FR\,I sources, the
lobe containing the main (brighter) jet is less depolarized than the
counter-jet lobe \citep{Morganti97b}. This can be explained if 
the main jet points toward the observer, suggesting that the
brightness asymmetry is caused by Doppler beaming \citep{Laing88}. The 
asymmetry decreases with distance from the nucleus \citep{Laing99},
implying that the jets must decelerate.  Self-consistent models of the
deceleration of relativistic jets by injection of matter lost from stars
or entrained from the surrounding galactic atmosphere have been calculated
by \citet{Bicknell94}, \citet{Komissarov94} and \citet{Bowman96}.

\citet[][hereafter LB]{LB} fit VLA images of the radio jets in the nearby
FR\,I radio galaxy 3C\,31 using a sophisticated model to reproduce the
observed total and polarized emission within 30\,arcsec of the nucleus,
where the jets are straight. They parameterized the
three-dimensional distributions of velocity, emissivity and magnetic-field
structure, calculated the brightness at each point within the jets in
Stokes $I$, $Q$ and $U$, accounting for the effects of relativistic
aberration, and integrated along the line of sight to reproduce the
expected distributions on the sky.  They concluded that the jets in 3C\,31
could be accurately modelled as intrinsically identical, axisymmetric,
antiparallel, decelerating, relativistic flows, with locally random but
anisotropic magnetic fields. Optimization of the model parameters placed
tight constraints on the geometry, velocity, emissivity and field
structure. \citet{LB2} used this velocity field, together with estimates
of the external pressure and density from {\sl Chandra} observations
\citep{Hardcastle02} in a conservation-law analysis based on that of
\citet{Bicknell94}. They showed that there are self-consistent solutions
for jet deceleration by injection of thermal matter and derived the
variations of pressure, density, Mach number and entrainment rate along
the jets. Finally, \citet{Laing04} developed models of adiabatic,
relativistic jets with velocity shear and applied them to 3C\,31. They
demonstrated that such models provide a reasonable description of the
emissivity and magnetic-field variations at large distances from the
nucleus but fail closer in, and inferred that significant reacceleration
of relativistic particles is required precisely where X-ray synchrotron
emission is observed \citep{Hardcastle02}.

In the present paper we apply a modified version of LB's model to the
jets of two FR\,I radio galaxies: B2\,0326+39 and B2\,1553+24. Our
principal aim is identical to that of LB: to estimate the
distributions of velocity, emissivity and magnetic-field structure
without introducing preconceptions about the (poorly understood)
internal physics. By studying different objects we hope to be able to
improve the range of physical scales we are able to probe, to identify
which intrinsic features are common to all FR\,I jets and which vary
from object to object and to assess the dependence of the jet
structure on power and environmental conditions.

Section \ref{obs} presents our new, deep, full-synthesis VLA images.  In
Section \ref{model}, we outline the model, emphasizing the
parameterizations of the geometry, velocity, emissivity and magnetic field
which differ from those used by LB. Section \ref{results} compares our
best-fitting models with the observed data in a variety of ways to show
the features that we are able to reproduce as well at those we cannot. In
Section \ref{physical}, we present the velocity, emissivity and field
structures of the best-fitting models. Section \ref{discuss} investigates
whether the magnetic-field structure and emissivity are consistent with
the assumptions of flux freezing and pure adiabatic energy loss and
examines the idea that the jets reaccelerate. We then
consider the appearance of the model for B2\,1553+24 at large angles to
the line of sight and the implications for the detectability of side-on
counterparts. Finally, we briefly compare the models for 3C\,31,
B2\,0326+39 and B2\,1553+24. Section \ref{ssfw} summarizes our conclusions
and outlines possible avenues for further work.

We adopt a Hubble constant, $H_0$ = 70\,$\rm{km\,s^{-1}\,Mpc^{-1}}$
throughout and define spectral index $\alpha$ in the sense $S(\nu) \propto
\nu^{-\alpha}$. We use the notation $p = (Q^2+U^2)^{1/2}/I$ for
the degree of linear polarization. 

%--------------------------- ****** DATA ****** -----------------------------%
\section{Observations and images}
\label{obs}
\subsection{Object selection}
\label{req}

Our models assume that the jets are intrinsically identical, antiparallel,
axisymmetric stationary flows, so any apparent differences must be due
entirely to relativistic effects. Some important selection criteria are
therefore imposed on our choice of source:
\begin{enumerate}
\item Only straight parts of the jets can be modelled and it is therefore
important that the first bends in both the main and counter-jets be as far
from the nucleus as possible.
\item The jet and counter-jet must leave the nucleus in strictly antiparallel
directions.
\item We derive the {\sl sidedness ratio} by dividing a total-intensity
image of the jets by a copy of itself rotated through 180$^\circ$ about
the nucleus. The ratio is in the sense main jet/counter-jet and cannot be
$<$1 over a significant area if the asymmetries are to be attributed
entirely to relativistic effects (we expect some local fluctuations due to
small-scale structure). This excludes a few objects, in particular those
where the counter-jet is wider than the main jet (e.g.\ B2\,0755+379;
\citealt{Bondi00}).
\item There cannot be any other radio features (e.g.\ lobes) that could be
confused with the emission from the jet. 
\item The jets must be bright enough to produce a significant polarized
signal from  the region to be modelled. This is required in order to
break the degeneracy between velocity and angle to the line of sight
which is inherent in the total intensity (Section~\ref{Overview}).
\end{enumerate}

We selected two suitable targets from the well studied sample of nearby radio
galaxies identified with sources in the B2 catalogue \citep{Colla75,Parma87}.

\subsection{B2\,0326+39}
\label{0326intro}

B2\,0326+39 is a typical FR\,I radio source associated with a bright
elliptical galaxy at $z=0.0243$ \citep{Miller02}. Its luminosity at
1.4\,GHz is $P_{1.4} = 2.2 \times 10^{24}$\,WHz$^{-1}$. 5-GHz images from the
WSRT \citep{Parma82} and VLA \citep{Bridle82} first resolved the two
fairly symmetrical, antiparallel jets. Comprehensive observations at
multiple frequencies and resolutions between 26 and 0.6\,arcsec were made by
\citet{Bridle91}, who showed that the main (western) jet is much
brighter than the counter-jet for the first 4 arcsec from the nucleus and
that the change from one-sided to two-sided structure is accompanied by a
flip in the polarization from parallel to perpendicular apparent
magnetic field.  The surrounding lobe emission is resolved out in images
with FWHM $\leq$ 2\,arcsec. The mean spectral index of the jets close to
the nucleus is $\alpha = 0.55$ between 1.4 and 5.0\,GHz \citep{Bridle91}.

{\sc ROSAT} PSPC observations \citep{Canosa99} reveal a point source with
a flux of 36\,nJy at 1\,keV and an extended atmosphere well fitted by a beta 
model with form factor $\beta_{atm} = 0.35$ and core radius $r_c = 30$\,kpc.

\subsection{B2\,1553+24}
\label{1553intro}

B2\,1553+24 is a more distant ($z = 0.0426$; \citealt{Colla75}) and less
luminous ($P_{1.4} = 4.7 \times 10^{23}$\,WHz$^{-1}$) FR\,I radio galaxy.
VLA observations at 1.4 and 4.8\,GHz \citep{Morganti87} revealed two
straight, well-collimated jets aligned NW -- SE with very little
surrounding lobe emission (see also \citealt{Stocke87};
\citealt{deRuiter93}).  Faint extended emission associated with the
brighter jet is visible on the NVSS, however \citep{NVSS}. The jets are
similar at large distances from the  bright compact nucleus but become
very asymmetric in the central 5\,arcsec. Their mean spectral index is
$\alpha = 0.60$ \citep{Morganti87}. The bright radio, optical and X-ray
core (\citealt{Laing99}; \citealt{Capetti00}; \citealt{Canosa99}), the
very large side-to-side asymmetries between the two jets close to the
nucleus \citep{Laing99} and the detection of optical emission from the
base of the main jet using HST \citep{Parma03} all indicate that the jets
are close to the line of sight.

\subsection{Observations and data reduction}

The observations were made using the VLA with a centre frequency of
8.46\,GHz and a 100-MHz bandwidth in all four standard configurations
for B2\,0326+39 and in A, B and C configurations for B2\,1553+24
(Table \ref{record}).  Either 3C\,48 or 3C\,286 was used as the
primary calibrator for amplitude and one of 3C\,138 or 3C\,286 was
observed to set the {\bf E}-vector position angle. The data for each
configuration were reduced separately in the {\sc AIPS} software
package using standard techniques of calibration, imaging and
self-calibration. The datasets were then concatenated using the {\sc
  AIPS} task {\sc DBCON} starting with the widest configurations (A
and B), first adjusting the core flux of the smaller-configuration
dataset to match that observed with the larger configuration and
imaged at matched resolution. A further iteration of phase
self-calibration was done after each combination. This procedure accounts for any
core flux-density variation between observations. {\sc CLEAN}ed images
were produced at three resolutions: full (FWHM 0.25 arcsec) and
tapered to two lower resolutions giving better signal-to-noise ratio
in the outer parts of the jets. $I$ images were also produced using
maximum-entropy deconvolution, first subtracting a point source at the
position of the core. After the deconvolution step, the point source
was added back in and the image convolved with the same truncated
Gaussian beam as used in the {\sc CLEAN}ed images. The maximum-entropy
and {\sc CLEAN}ed images were consistent apart from some low-level
artefacts clearly associated with the latter. We therefore show only
the maximum-entropy $I$ images. The $Q$ and $U$ images were {\sc
  CLEAN}ed. Noise levels for $I$ and $Q/U$ images are given in Table
\ref{noise} and are close to theoretical limits.  All of the displayed
polarization vectors have been rotated by 90$^\circ$ from the
direction of the {\bf E}-vector at 8.4\,GHz and therefore represent
apparent {\em magnetic} field. No high-resolution, multifrequency
studies of Faraday rotation have yet been made for the two sources, so
we have not corrected for this effect. The estimates of rotation
measure from two-frequency data given by \citet{Bridle91} and
\citet{Morganti87,Morganti97a} and direct comparison of their
higher-frequency images with ours implies that the magnetic field
directions inferred from our data are in error by $<$2$^\circ$ due to
Faraday rotation. No corrections for Ricean bias \citep{WK} have been
made in deriving the images of degree of polarization, $p$, presented
in this paper, but they have all been blanked at low signal-to-noise
ratio as indicated in the figure captions, so the bias is
negligible. In any case, the fitting procedure described below (which
uses the $Q$ and $U$ images directly) is unaffected.

\begin{center}
\begin{table}
\caption{Record of VLA observations}
\begin{center}
\begin{tabular}{lclc}
\hline
Object      & Configuration &    Date     & Integration \\
            &               &             & time (min)  \\
\hline
B2\,0326+39 &       A       & 1999 Sep 18 &     604     \\
            &       B       & 1999 Dec 19 &     604     \\
            &       C       & 2000 May 14 &     131     \\
            &       D       & 2000 Aug 17 &      41     \\
&&&\\
B2\,1553+24 &       A       & 1999 Sep 18 &     603     \\
            &       B       & 1999 Dec 19 &     321     \\
            &       C       & 2000 May 14 &     107     \\
\hline
\end{tabular}
\end{center}
\label{record}
\end{table}
\end{center}

\begin{center}
\begin{table}
\caption{Image resolutions and noise levels. $\sigma_I$ is the
  off-source noise level on the $I$ image; $\sigma_P$ the average of
  the noise levels for $Q$ and $U$.
\label{noise}}
\begin{tabular}{llcc}
\hline
Object    & FWHM  &\multicolumn{2}{c|}{rms noise level} \\
          &  (arcsec)  &\multicolumn{2}{c|}{($\mu$Jy / beam area)} \\
          &            &$\sigma_I$&$\sigma_P$ \\
\hline
B2\,0326+39 &   1.50   & 7.6 & 6.2 \\      
	    &   0.50   & 5.7 & 6.0 \\
	    &   0.25   & 5.8 & 6.8 \\
            
&&&\\
B2\,1553+24 &   1.50   & 5.2 & 5.8 \\
	    &   0.75   & 5.7 & 5.7 \\
	    &   0.25   & 6.0 & 6.5 \\

\hline
\end{tabular}
\end{table}
\end{center}

\begin{center}
\begin{table}
\caption{Core flux densities and positions}
\label{core}
\begin{tabular}{lccc}
\hline
Object      & Core flux & \multicolumn{2}{c|}{Core Position (J2000)} \\
	    &   (mJy)   &    RA     & Dec         \\
\hline
B2\,0326+39 &  71.4   & $03^h 29^m 23.887^s$ & $39^\circ 47' 31.96''$ \\
B2\,1553+24 &  63.5   & $15^h 56^m 03.912^s$ & $24^\circ 26' 52.92''$ \\
\hline
\end{tabular}
\end{table}
\end{center}

\subsection{Source description}

The positions and 8.4-GHz flux densities of the cores found from our
A-configuration images are given in Table \ref{core}.

\subsubsection{B2\,0326+39}
\label{0326}

\begin{figure*}
\includegraphics[width=17cm]{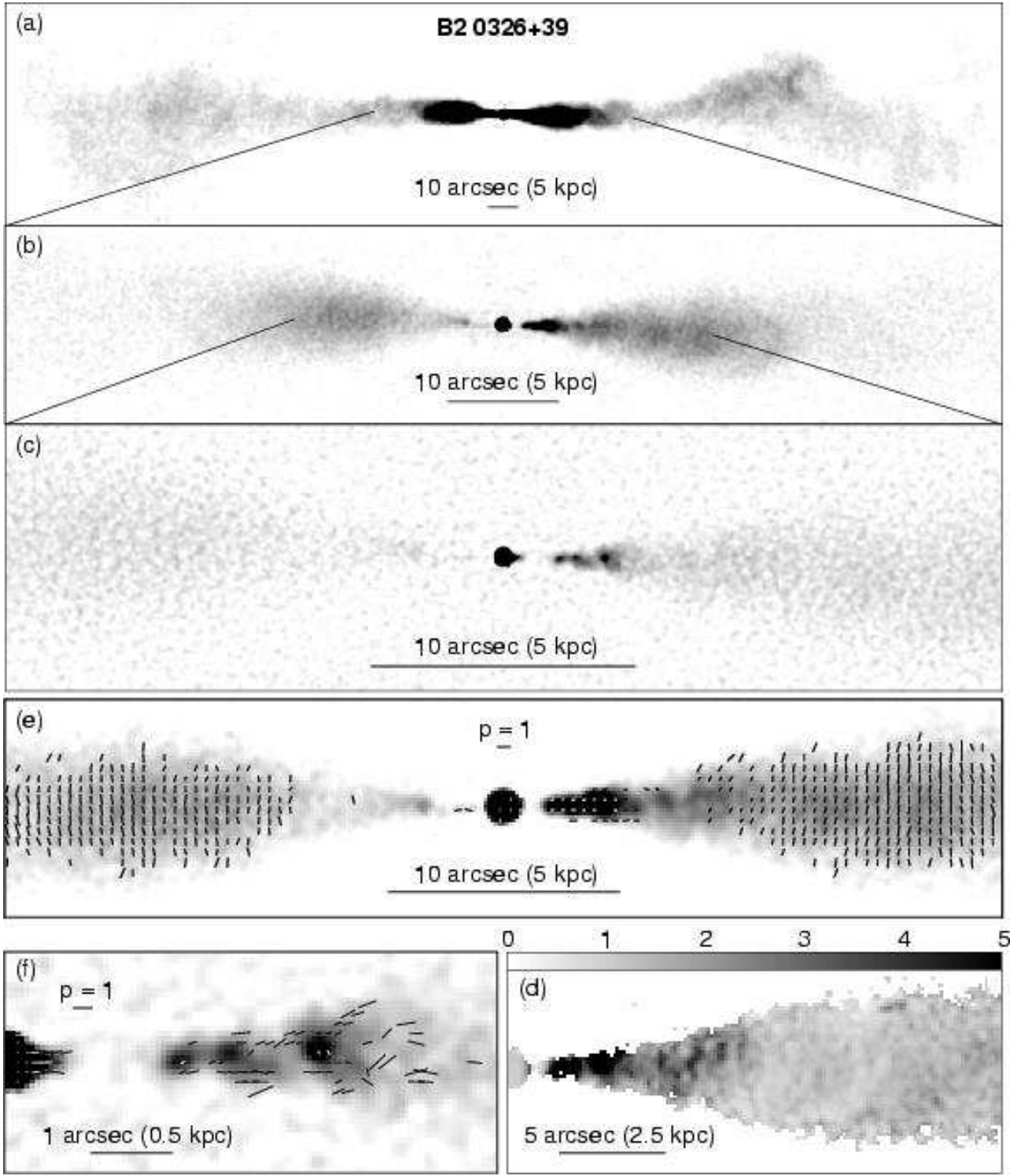}
\caption{Montage showing the large-scale structure and jets of B2\,0326+39
at 8.4\,GHz. Panels (a) -- (c) have North at the top; in the remaining
panels, the images have been rotated slightly so that the jet axis is
horizontal.  (a) 5.7\,arcmin (170\,kpc) East-West field at 1.5\,arcsec
(0.75\,kpc) resolution. (b) 1.5\,arcmin (45\,kpc) East-West field at
0.50\,arcsec (0.25\,kpc) resolution. (c) 38\,arcsec (19\,kpc) East-West
field at 0.25\,arcsec (0.13\,kpc) resolution.  (d) The jet/counter-jet
intensity ratio (defined as in Section~\ref{req}) for the straight region
of the jet within 22\,arcsec of the nucleus at 0.50\,arcsec (0.25\,kpc)
resolution. The grey-scale range (0 -- 5) is indicated by the labelled
bar. (e) Grey-scale of total intensity with superimposed polarization
vectors for the straight region of
the jets within 22\,arcsec of the nucleus at 0.50\,arcsec (0.25\,kpc)
resolution.  The lengths of the vectors are proportional to the degree of
polarization $p$ and their directions are those of the apparent magnetic
field. (f) As in panel (e), but showing the inner 6\,arcsec of the western
jet at 0.25\,arcsec (0.13\,kpc) resolution.
\label{fig:0326.montage}}
\end{figure*}

\begin{figure*}
\includegraphics[width=16.5cm]{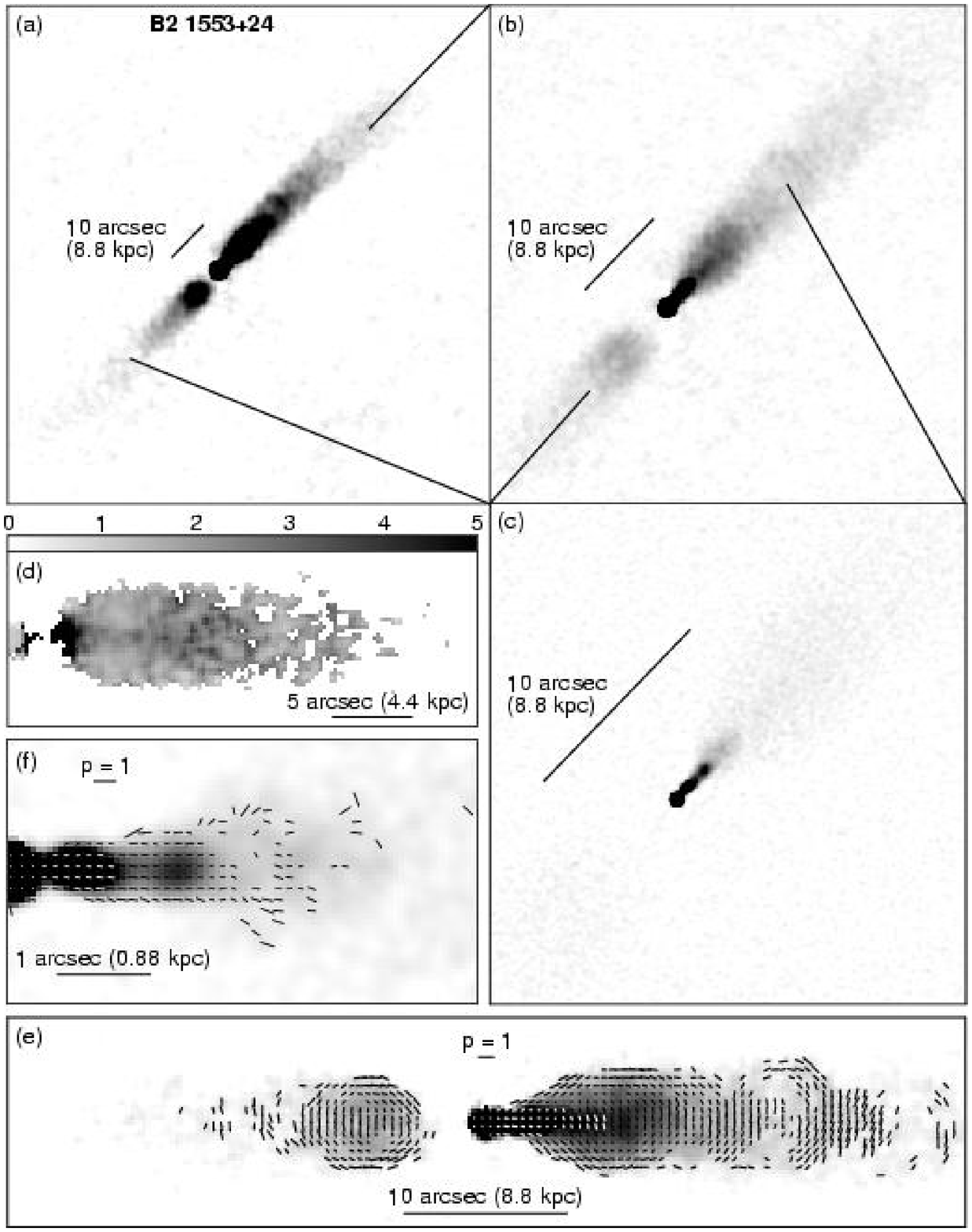}
\caption{Montage showing the jets of B2\,1553+24 at 8.4\,GHz.  Panels
  (a) -- (c) show grey-scales of total intensity and have North at the
  top; the remaining images have been rotated so that the jet axis is
  horizontal.  (a) 111\,arcsec (98\,kpc) East-West field at
  1.5\,arcsec (1.3\,kpc) resolution.  (b) 53\,arcsec (47\,kpc)
  East-West field at 0.75\,arcsec (0.66\,kpc) resolution. (c)
  22.5\,arcsec (20\,kpc) East-West field at 0.25\,arcsec (0.22\,kpc)
  resolution. (d) The jet/counter-jet intensity ratio for the inner
  (straight) section within 30\,arcsec (26\,kpc) of the nucleus at
  0.75\,arcsec (0.66\,kpc) resolution. The labelled wedge indicates
  the grey-scale range. (e) A grey-scale of the total intensity with
  superposed polarization vectors, showing 30\,arcsec from the nucleus
  of each jet at 0.75\,arcsec (0.66\,kpc) resolution. The superposed
  vectors indicate the degree of polarization $p$ (on a scale given by
  the labelled bar) and the direction of the apparent magnetic
  field. (f) As in panel (e), but for the inner 5\,arcsec of the main
  jet at 0.25\,arcsec (0.22\,kpc) resolution.
\label{fig:1553.montage}}
\end{figure*}

Figs~\ref{fig:0326.montage}(a) -- (c) show the jets of B2\,0326+39 at three
different resolutions. In panel (a), the jets can be seen over their full
extent. They are initially straight and antiparallel, although both bend
slightly (by $\approx$3$^\circ$) about 25\,arcsec from the nucleus. At a distance
of $\approx$22\,arcsec the ridge line of the main jet starts to oscillate
slightly. Panel (b) shows the inner 45\,arcsec of both jets at 0.5-arcsec
resolution: the jets appear very different from each other within
$\approx$10\,arcsec of the nucleus but become almost indistinguishable further
out. This is also illustrated in Fig \ref{fig:0326.montage}(d), which shows the
ratio of the main and counter-jet intensities at the same resolution. Panel (c)
shows the jets at our highest resolution, 0.25\,arcsec. The brightest region of the
main jet is resolved into three knots embedded in more extended emission. The
nucleus has an extension pointing in the direction of the main jet whose
brightness drops rapidly with distance from the nucleus; it can only be seen out
to $\approx$1\,arcsec. No emission is visible at 0.25-arcsec resolution between
this point and the first bright knot of emission $\approx$2\,arcsec from the
nucleus.

The polarization structure of B2\,0326+39 is shown at resolutions of 0.50
and 0.25\,arcsec in Figs~\ref{fig:0326.montage}(e) and (f),
respectively. As is usual in FR\,I objects \citep{Bridle84}, the main jet
close to the nucleus is polarized with an apparent field parallel to its
axis. Further out in the main jet and in all parts of counter-jet with
sufficient signal-to-noise ratio, the apparent field is perpendicular to
the axis. In this object, unlike 3C\,31 and B2\,1553+24, no parallel
field edge is seen.

We model the jets within 22\,arcsec of the nucleus, where their ridge lines 
are straight.

\subsubsection{B2\,1553+24}
\label{1553}

Figs~\ref{fig:1553.montage}(a) -- (c) show the jets of B2\,1553+24 at
three different resolutions: 1.5, 0.75 and 0.25\,arcsec. The jets are very
straight with only a slight bend in the main jet about 45 arcsec from the
nucleus. Both jets gets steadily fainter further out, and are
lost in the noise approximately 65 and 45\,arcsec from the nucleus for the
main and counter-jets respectively in our lowest (1.5\,arcsec) resolution
image. The counter-jet is invisible within 3\,arcsec of the nucleus; in contrast, 
the main jet is very bright there. Further out, the counter-jet
brightens, reaching a maximum at 7\,arcsec from the nucleus. From this
point outwards, the main and
counter-jets appear essential identical, but  with the main jet
approximately twice as bright. This is also illustrated 
in the image of jet/counter-jet intensity ratio 
(Fig.~\ref{fig:1553.montage}d). Initially (within 1.5\,arcsec) the main
jet is well collimated, as are both jets at distances $\geq$7\,arcsec.
Both jets flare dramatically between these regions of good collimation.

The polarization structure of the jets in B2\,1553+24 is shown in
Figs~\ref{fig:1553.montage}(e) and (f) at resolutions of 0.75 and
0.25\,arcsec respectively. The bright 4\,arcsec of the main jet closest to
the nucleus has an apparent magnetic field aligned along the jet
axis. Further out, the field becomes perpendicular to the axis close
to the centre of the jet and parallel to the outer isophotes near its
edges.

We model the jets out to 30\,arcsec from the nucleus. This limit is set
not by changes in jet direction, as in 3C\,31 and B2\,0326+39, but by
signal-to-noise, particularly in polarization. 

%-------------------------- ****** MODEL ****** ----------------------------%
\section{The model}
\label{model}

\subsection{Assumptions}
\label{Model-assumptions}

Our fundamental assumptions are those of LB:
\begin{enumerate}
\item The jets may be modelled as antiparallel, axisymmetric, stationary, 
laminar flows.  
\item They contain relativistic particles with an energy spectrum $n(E)dE
= n_0 E^{-(2\alpha+1)}dE$ (corresponding to a frequency spectral index
$\alpha$) with an isotropic pitch-angle distribution. The maximum degree
of linear polarization is then $p_0 = (3\alpha+3)/(3\alpha+5)$. We assume
that the 1.4 -- 5\,GHz spectral indices given in Sections~\ref{0326} and
\ref{1553} also apply at 8.4\,GHz.
\item The magnetic field is tangled on small scales, but anisotropic
(the reasons for taking the field to be of this form are discussed by
  \citealt{Laing81}, \citealt{BBR} and LB).
\end{enumerate}
We define $\beta = v/c$, where $v$ is the flow velocity. $\Gamma =
(1-\beta^2)^{-1/2}$ is the bulk Lorentz factor. 

\subsection{Overview}
\label{Overview}

We aim to determine the velocity and angle to the line of sight
independently by comparison of emission from the main and
counter-jets and the modelling of linear polarization is essential to
our technique. In order to illustrate this point, we consider the
simple example of cylindrical, antiparallel jets with constant velocity
$\beta c$ at an angle $\theta$ to the line of sight.  If we
consider only total intensity, the ratio of the flux densities per
unit length for the main and counter-jets, $I_{\rm j}/I_{\rm cj}$,
does not uniquely determine the velocity:
\begin{eqnarray*}
\frac{I_{\rm j}}{I_{\rm cj}} & = & \left( \frac
{1+\beta\cos\theta}{1-\beta\cos\theta}\right)^{2+\alpha} \\ 
\end{eqnarray*}
for emission which is isotropic in the frame of the jet flow. In order
to break the degeneracy between $\beta$ and $\theta$, we use the
linear polarization.  The relation between the angles to the line of
sight in the rest frame of the flow, $\theta^\prime$ and in the
observed frame, $\theta$, is:
\begin{eqnarray*} 
\sin\theta^\prime_{\rm j} & = & [\Gamma(1-\beta\cos\theta)]^{-1}\sin\theta
\makebox{~~~~~(main jet)} \\
\sin\theta^\prime_{\rm cj} & = & [\Gamma(1+\beta\cos\theta)]^{-1}\sin\theta
\makebox{~~(counter-jet)} \\
\end{eqnarray*}
The observed polarization is in general a function of $\theta^\prime$
if the field is anisotropic.  If we know the field structure a priori,
then we can solve explicitly for $\beta$ and $\theta$. We take the
example of a field which is disordered on small scales but confined to
a plane perpendicular to the jet, with equal rms along any direction
in the plane. In this case, there is no variation of the degree or
direction of polarization across the jet. For $\alpha = 1$, the total
and polarized flux densities per unit length in the emitted frame are:
\begin{eqnarray*}
I^\prime & = & K(1+\cos^2\theta^\prime) \\
P^\prime & = & p_0 K\sin^2\theta^\prime \\
\end{eqnarray*}
where $K$ is a constant \citep{Laing80,Laing81}.  The ratios of
observed total and polarized intensity for the jet and counter-jet
are:
\begin{eqnarray*}
\frac{I_{\rm j}}{I_{\rm cj}} & = & \left[ \frac{2 -
    [\Gamma(1-\beta\cos\theta)]^{-2} \sin^2\theta}{2 -
    [\Gamma(1+\beta\cos\theta)]^{-2} \sin^2\theta} \right]^3
    \\
\frac{P_{\rm j}}{P_{\rm cj}} & = &
    \left(\frac{1+\beta\cos\theta}{1-\beta\cos\theta}\right)^5
    \\
\end{eqnarray*}
These equations can be solved numerically for $\beta$ and $\theta$
(cf. \citealt{Bondi00}). [The specific field configuration is a
reasonable approximation for the outer regions of the jets in
B2\,0326+39 (Section~\ref{0326magnetic})].

In general, we must fit the field configuration and therefore need to
introduce two additional parameters to describe its anisotropy
(Section~\ref{bfield}). These also determine the variation of
polarization transverse to the jet axis, and can be estimated
independently of the velocity and angle if the jets are well resolved
in this direction.
 
Following LB, we model the jets using simple parameterized forms for
the jet geometry, velocity, emissivity and magnetic-field
structure. We find that the functional forms used by LB to model
variations along the jets are unable to fit the new data. In order to
produce models that fit the data accurately, we are forced to decouple
the parameterizations of velocity, emissivity and magnetic field from
the three regions defined by the geometry in LB. This provides much
more freedom in our models, but also allows simpler functional forms
to be used.

LB examined two forms for transverse variations of velocity and emissivity:
{\em spine/shear-layer (SSL)} and {\em Gaussian}. The former gave a
slightly better fit to the data for 3C\,31, at the cost of introducing
extra parameters.  We find that the off-axis variations of velocity and
emissivity are less well constrained in our objects and that the difference
between SSL and Gaussian models is insignificant. In what follows, we use
only the simpler Gaussian models.

\subsection{Geometry \& Coordinate Systems}
\label{Geometry}

\begin{figure*}
\includegraphics[width=16.5cm]{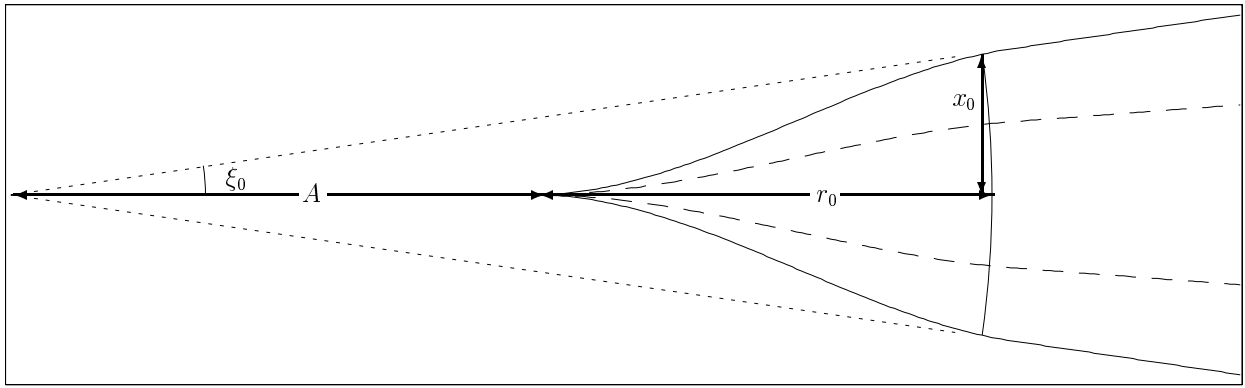}
\caption{Geometry of the model. The jet edge ($s = 1$ streamline) is
indicated by the solid line and the $s = 0.5$ streamline by the dashed
line. The arc shows the spherical boundary between the flaring and outer
regions, centred at the vertex of the outer region. The three quantities
used to define the jet geometry: 
$r_0$, the distance of the outer boundary from the nucleus; $x_0$, the
width of the jet at the outer boundary and $\xi_0$, the opening angle of
conical outer region are marked.
\label{fig:geom}}
\end{figure*}

The geometry of the model jet is sketched in Fig.~\ref{fig:geom}. It is
divided into two parts: a {\em flaring region} where the jet first expands
rapidly and then recollimates and an {\em outer region} of constant
opening angle. The mathematical description of the edge of the jet is as
given by LB, but with two important changes.  First, the conical outer
region is not centred on the nucleus but at a distance $A$ behind the
nucleus along the jet axis, allowing the jet to recollimate much more than
is seen in 3C\,31 (Fig.~\ref{fig:geom}). This modification is required to
describe many FR\,I jets, including both B2\,0326+39 and
B2\,1553+24. Second, we find no evidence for separate conical inner
regions in the B2 sources (the jets are very narrow and poorly resolved
close to the nuclei, so their collimation behaviour is poorly constrained
there).  We therefore extend the flaring region to the base of the jet. An
inner region analogous to that used by LB to model 3C\,31 is defined
purely by the emissivity profile (Section \ref{Emissivity}).The distance
of the outermost isophote from the jet axis in the flaring region is fit
by a cubic polynomial, as in LB, but both it and its derivative are
constrained to be zero at the nucleus ($\xi_1 = 0$ and $r_1 = 0$ in the
notation of LB).

Four parameters define the geometry of the model.  Of these, the angle
to the line of sight, $\theta$, the opening angle of the conical outer
region, $\xi_0$, and the distance along the axis from the nucleus to
the boundary between flaring and outer regions, $r_0$, are defined
essentially as in LB.  Because the vertex of the outer region is no
longer at the nucleus, however, the spherical boundary surface is
centred on the vertex and has radius $r_0 + A$ (this reduces to
the case considered by LB if $A = 0$). The fourth defining parameter is
$x_0$, the radius of the jet edge at the outer boundary
(Fig.~\ref{fig:geom}). We choose to optimize $x_0$ rather than $A$ to
avoid cross-coupling between variables: $x_0$ is determined by the
position of the outer isophote at the boundary and $\xi_0$ by the
expansion of the jet in the outer region, whereas $A$ affects both
quantities.  The variables are related by $(r_0 + A)\sin\xi_0 = x_0$.

The velocity, emissivity and magnetic-field parameterizations defined
below are expressed in a streamline coordinate system $(\rho,s)$ where the
streamline index $s$ is constant for a given streamline and $\rho$ increases
monotonically with distance along it. The distance of a streamline from
the jet axis is:
\begin{eqnarray*}
x(z,s) & = & a_2(s) z^2 + a_3(s) z^3 \makebox{~~~~(flaring region)} \\
x(z,s) & = & (z+A)\tan (\xi_0 s) \makebox{~~~~~~(outer region)}\\
\end{eqnarray*}
where $z$ is the distance from the nucleus along the axis. In the
outer region, $s = \xi/\xi_0$, where $\xi$ is the angle between the
flow vector and the jet axis. For the flaring region, $a_2(s)$ and
$a_3(s)$ are defined by the conditions that $x(z,s)$ and its
derivative with respect to $z$, $x^\prime(z,s)$, are continuous at the
boundary between the two regions.  $s$ therefore varies from 0 on-axis
to 1 at the jet edge; the $s = 0.5$ streamline is plotted in
Fig.~\ref{fig:geom}. The form of $x(z,s)$ in the flaring region is
that used by LB with the simplifying condition $x(0,s) = x^\prime(0,s)
= 0$. Parameterization of physical quantities in terms of the distance
from the nucleus along a streamline creates undue complexity because
of the cubic form of $x(z,s)$. Instead, we use a coordinate:
\begin{eqnarray*}
\rho & = & \frac{zr_0}{(r_0 + A) \cos (\xi_0s) - A} \hspace{1.0cm}
\rho < r_{0} \\ \rho & = & \frac{z + A}{\cos (\xi_0s)} - A
\hspace{2.25cm} \rho \ge r_{0} \\
\end{eqnarray*}
$\rho$ increases monotonically along a streamline from 0 at the nucleus
and each streamline crosses the boundary into the outer region at $\rho =
r_0$. The coordinate system used by LB is the special case with $A = 0$.

\subsection{Velocity field}
\label{Velocity}

The variations in sidedness radio along the jets in both sources
(Figs~\ref{fig:0326.montage}d, \ref{fig:1553.montage}d) suggest that the
initial velocity needs to be high in order to account for the asymmetry as
a consequence of Doppler boosting.  Further from the nucleus, 
there is a transition to a regime where the sidedness ratios
are low and approximately uniform. In B2\,1553+24, the  ratio
remains constant, but significantly $>$1 for all distances $>$7\,arcsec
from the nucleus, indicating a mildly relativistic but constant velocity.
Two changes to the form of the velocity profile used by LB to
describe 3C\,31 are required:
\begin{enumerate}
\item In the flaring region, the function used by LB describes an initial
fast section and abrupt deceleration, just as needed for the B2 sources, but
the deceleration is forced to occur immediately before the jet recollimates. This
does not appear to be general, so we need a functional form which is
qualitatively similar to that used by LB, but which allows the location
and length of the rapid deceleration region to be varied.
\item The jet/counter-jet sidedness ratio in the outer region of 3C\,31
continues to decline slowly with distance from the nucleus, so LB used an
exponential form for the velocity profile there. This will not fit the
large regions of constant sidedness ratio in the B2 sources, especially
in B2\,1553+24.
\end{enumerate} 
We therefore divide the on-axis velocity profile into three regions: (a)
approximately constant, with a high velocity close to the nucleus; (b) a
linear fall-off and (c) roughly constant, but with a low velocity at large
distances. Example profiles are given later 
(Figs~\ref{fig:0326profiles}b and \ref{fig:1553profiles}b).  The profile
is defined by four free parameters: the distances of the two boundaries
separating the three regions, $\rho_{\beta_1}$ and $\rho_{\beta_0}$, and
the characteristic inner and outer velocities $\beta_1$ and $\beta_0$.

The velocity along any off-axis streamline is
calculated using the same expressions but with inner and outer velocities
$\beta_1\exp(-s^2\ln v_1)$ and $\beta_0\exp(-s^2\ln v_0)$, respectively,
i.e.\ with a truncated Gaussian transverse profile falling to fractional
velocities $v_1$ and $v_0$ at the edge of the jet in the inner and outer
regions, respectively.  This differs slightly from the separable function
used by LB, but gives essentially identical transverse profiles.

The full functional forms for the velocity field $\beta(\rho,s)$ are given
in Table~\ref{tab:param}. [Note that the exponential terms are introduced
purely to maintain continuity in the acceleration near the boundaries and
their precise form has no significant effect on the brightness
distribution].  The constants $c_1$, $c_2$, $c_3$ and $c_4$ are defined by
the values of the free parameters and the conditions that the velocity and
acceleration are continuous at the two boundaries.

\begin{table*}

\caption{Functional forms of the velocity $\beta$, emissivity
$\epsilon$, radial/toroidal and longitudinal/toroidal magnetic-field
ratios $j$ and $k$ in the streamline coordinate system
$(\rho,s)$. Column 4 lists the parameters which may be optimized, for
comparison with Table~\ref{Params}.}
\begin{center}
\begin{tabular}{llcl}
\hline
&&&\\
Quantity & Functional form & Range & Free parameters \\
&&&\\
\hline
&&&\\
\multicolumn{4}{c}{Velocity field}\\
&&&\\
$\beta(\rho,s)$ &  $\beta_{1} - \left[\frac{\beta_{1}\exp(-s^2\ln v_1) - \beta_{0}\exp(-s^2\ln v_0)}{10}\right]\exp[c_{1}(\rho - \rho_{\rm v_1})]$ 
& $\rho < \rho_{\rm v_1}$ & Distances $\rho_{\rm v_1}$, $\rho_{\rm
  v_0}$ \\
&&&\\
& $c_{2} + c_{3}\rho$ 
& $\rho_{\rm v_1} \le \rho \le \rho_{\rm v_0}$ & Velocities $\beta_1$,
$\beta_0$\\
&&&\\
& $\beta_{1} - \left[\frac{\beta_{1}\exp(-s^2\ln v_1) - \beta_{0}\exp(-s^2\ln v_0)}{10}\right]\exp[c_{4}(\rho_{\rm v_0} - \rho)]$ 
& $\rho > \rho_{\rm v_0}$ & Fractional edge velocities $v_1$, $v_0$\\
&&&\\
&&&\\
\multicolumn{4}{c}{Emissivity}\\
&&&\\
$\epsilon(\rho,s)$ & $~g\left(\frac{\rho}{\rho_{\rm e_1}}\right)^{-E_1}$ 
& $\rho \le \rho_{\rm e_1}$ & Distances $\rho_{\rm e_1}$,$\rho_{\rm
  e_2}$, $\rho_{\rm e_3}$,$\rho_{\rm e_4}$ \\
&&&\\
&~~~$\left(\frac{\rho}{\rho_{\rm e_1}}\right)^{-E_2}
\exp\left[-s^2\ln\left(e_1 + (e_0 - e_1)\left(\frac{\rho - \rho_{\rm e_1}}{\rho_{\rm e_2} - \rho_{\rm e_1}}\right)\right)\right]$ 
& $\rho_{\rm e_1} < \rho \le \rho_{\rm e_2}$ & Indices $E_1$, $E_2$, $E_3$, $E_4$, $E_5$  \\
&&&\\
& $d_1\left(\frac{\rho}{\rho_{\rm e_2}}\right)^{-E_3}
\exp(-s^2\ln e_0)$ 
& $\rho_{\rm e_2} < \rho \le \rho_{\rm e_3}$ & Fractional increase $g$\\
&&&\\
&$d_2 \left(\frac{\rho}{\rho_{\rm e_3}}\right)^{-E_4} \exp(-s^2\ln e_0)$ 
& $\rho_{\rm e_3} < \rho \le \rho_{\rm e_4}$ & \\
&&&\\
&$d_3 \left(\frac{\rho}{\rho_{\rm e_4}}\right)^{-E_5} \exp(-s^2\ln e_0)$ 
& $\rho > \rho_{\rm e_4}$ &\\
&&&\\
&&&\\
\multicolumn{4}{c}{Radial/toroidal field ratio}\\
&&&\\
$j(\rho,s)$ &  $j_{1}$ 
& $\rho \le \rho_{\rm B_1}$ & Distances $\rho_{\rm B_1}$, $\rho_{\rm B_0}$\\
&&&\\
&$j_{1} + (j_{0} - j_{1})\left(\frac{\rho - \rho_{\rm B_1}} {\rho_{\rm B_0} - \rho_{\rm B_1}}\right)$ 
& $\rho_{\rm B_1} < \rho < \rho_{\rm B_0}$ & Ratios $j_1$, $j_0$\\
&&&\\
&$j_{0}$ 
& $\rho \ge \rho_{\rm B_0}$ &\\
&&&\\
&&&\\
\multicolumn{4}{c}{Longitudinal/toroidal field ratio}\\
&&&\\
$k(\rho,s)$ & $k_{1}$ 
& $\rho \le \rho_{\rm B_1}$ & Ratios $k_1$, $k_0$ \\
&&&\\
& $k_{1} + (k_{0} - k_{1})\left(\frac{\rho - \rho_{\rm B_1}} {\rho_{\rm B_0} - \rho_{\rm B_1}}\right)$ 
& $\rho_{\rm B_1} < \rho < \rho_{\rm B_0}$ &\\
&&&\\
& $k_{0}$ 
& $\rho \ge \rho_{\rm B_0}$ & \\
&&&\\
\hline
\end{tabular}
\end{center}
\label{tab:param}
\end{table*}

\subsection{Magnetic Field Structure}
\label{bfield}

We define the rms components of the magnetic field to be
$\langle B_l^2\rangle^{1/2}$ (longitudinal, parallel to a streamline),
$\langle B_r^2\rangle^{1/2}$(radial, orthogonal to the streamline and outwards from
the jet axis) and $\langle B_t^2\rangle^{1/2}$ (toroidal, orthogonal to the
streamline in an azimuthal direction).  The magnetic-field structure
is parameterized, as in LB, by the ratio of rms radial/toroidal field,
$j(\rho,s) = \langle B_r^2\rangle^{1/2}/\langle B_t^2\rangle^{1/2}$ and the
longitudinal/toroidal ratio $k(\rho,s)
=\langle B_l^2\rangle^{1/2}/\langle B_t^2\rangle^{1/2}$. The main difference between our model
and that of LB is again that the characteristic distances are defined
independently from those for other quantities. Two fiducial distances,
$\rho_{\rm B_1}$ and $\rho_{\rm B_0}$, are used. For $\rho < \rho_{\rm B_1}$ and
$\rho > \rho_{\rm B_0}$, the field ratios have constant values, with
linear interpolation between them for $\rho_{\rm B_1} \leq \rho \leq
\rho_{\rm B_0}$.  We find no evidence for any transverse structure in
either of the field ratios. In 3C\,31, there are significant
variations in the radial/toroidal field ratio between the centre and
edge of the jet, and the model of LB is therefore more complex than
that used here.  The functional forms assumed for the field ratios are
again given in Table~\ref{tab:param}.

\subsection{Emissivity}
\label{Emissivity}

We write the proper emissivity as $\epsilon(\rho,s) f(\rho, s)$, where
$\epsilon$ is the emissivity in $I$ for a magnetic field $B = \langle B_l^2 +
B_r^2 + B_t^2\rangle^{1/2}$ perpendicular to the line of sight. $f$
depends on field geometry: for $I$,
$0 \leq f \leq 1$ and for $Q$ and $U$ $-p_0 \leq f \leq +p_0$.
Following LB, we refer to $\epsilon$, loosely, as `the
emissivity'. For a given spectral index, it is a
function only of the rms total magnetic field and the normalizing
constant of the particle energy distribution, $\epsilon \propto n_0
B^{1+\alpha}$.

The description of the emissivity is similar to that given by LB: the
on-axis profile is divided into distinct regions, each with a power-law
profile.  As for the velocity, however, the boundaries between regions are
defined without reference to the geometry. In order to differentiate
between significant changes in the emissivity profile and small-scale,
stochastic features (which will occur at different places in the main and
counter-jets), we require evidence that the same emissivity structure is
present in both jets and that it extends over many beam areas.  Five
regions are required (although only four are used for each source)
compared with the three used by LB. The reasons for the increase in the
number of regions are:
\begin{enumerate}
\item In B2\,0326+39 the bright region of the main jet between 2 and
5\,arcsec from the nucleus (Fig.~\ref{fig:0326.montage}b) is composed of
three knots of emission of approximately equal brightness with a sharp
falloff at each end; there is a much fainter equivalent in the
counter-jet. Two regions are therefore required rather than the one used
in LB: a relatively flat section averaging the intensity of the knots and
a steep section modelling the rapid falloff at the end of the last
knot. The rapid increase in brightness at the innermost knot is modelled
using a discontinuity in the emissivity profile as in LB.
\item A further region is required to model the outer parts of both jets in
B2\,1553+24 (we will show that this is on a much larger physical scale
than the entire modelled regions of 3C\,31 and B2\,0326+39; it is
therefore reasonable that it shows different emissivity behaviour).
\end{enumerate}
The emissivity is continuous across all of the
boundaries except the innermost ($\rho = \rho_{\rm e_1}$), where a jump by
a factor of $g$ is allowed. The part of the jet with $\rho <  \rho_{\rm
e_1}$ is analogous to the inner region defined by geometry, emissivity and
velocity for 3C\,31 (LB).

Off-axis, the profile is multiplied by a factor
$\exp[-s^2\ln\bar{e}(\rho)]$, as in LB, so that  $\bar{e}(\rho)$ is the fractional
value of the emissivity at the jet edge.  $\bar{e}(\rho)$ has a constant
value $e_0$ for $\rho > \rho_{\rm e_2}$ and varies linearly through region
2 from $e_1$ at $\rho_{\rm e_1}$ to $e_0$ at $\rho_{\rm e_2}$. In the
inner region the jet is too narrow to constrain any transverse profile and
we set $\bar{e}(\rho) = 1$.

The full description of the emissivity distribution $\epsilon(\rho,s)$ is
given in Table~\ref{tab:param}. 

\subsection{Modelling Procedure}

The modelling procedure is that described by LB, with one small
refinement. In order for the numerical integrations to converge, it is
necessary to determine appropriate limits, which are where the line of
sight intersects the jet surface. If $\theta$ is small, a line of sight
may cross the jet boundary more than twice, for example in passing through
both the inner part of the jet and the shoulder where the jet
recollimates. This possibility is required for B2\,1553+24 and is now
included.

\subsection{Fitting and optimization}
\label{Fit-opt}

We use $\chi^2$ as a measure the goodness of fit. The noise
level for each resolution and Stokes parameter is estimated as
$1/\sqrt{2}$ times the rms difference between the image and a copy of
itself reflected across the jet axis, as in LB. This quantity is
always significantly larger than the off-source noise level. It is a
measure of the local deviation from mirror symmetry, which limits our
ability to fit the data using an axisymmetric model, and includes
a contribution from any residual deconvolution errors. It is a lower limit to
the correct noise level, as it excludes any mirror-symmetric
component. We therefore expect $\chi^2$ to be {\em overestimated}.

$\chi^2$ is calculated using the high-resolution (0.25\,arcsec) images
where they have adequate signal-to-noise ratio (i.e.\ close to the
nucleus) and the lower-resolution images elsewhere. $\chi^2$ values
for $I$, $Q$ and $U$ are summed. In a slight modification to the algorithm
used by LB, the sizes of the regions over which $\chi^2$ is
evaluated from the high-resolution data may be different for total
intensity and linear polarization.  For B2\,0326+39 we fit to the
0.25-arcsec $I$ within 4.5\,arcsec, further out, we use the 0.5-arcsec
image; for the $Q$ and $U$ images only the 0.5-arcsec images are used. 
For B2\,1553+24 we fit the inner 4\,arcsec of the 0.25-arcsec
$I$, $Q$ and $U$ images and the 0.75-arcsec images at larger
distances. The sampling grid is chosen so that all estimates of
$\chi^2$ are independent and the core is always excluded.

An initial starting model is generated by adjusting a number of key
parameters by hand to match recognizable features of the jet.  Their
values are then fixed and the remaining parameters are optimized using
the downhill simplex method of Nelder \& Mead implemented using
Numerical Recipes routines \citep{NR} to minimize $\chi^2$.

\begin{table}
\caption{The number of independent points and the reduced $\chi^2$ values
  for our final optimized models.
\label{tab:chisq}}
\begin{tabular}{lcc}
\hline
Object      & Independent& Reduced $\chi^2$ \\
            & points     &                  \\
&\\
B2\,0326+39 &   3614   &      1.32        \\      
&\\
B2\,1553+24 &   3722   &      1.13        \\
\hline
\end{tabular}
\end{table}

\begin{figure*}
\includegraphics[width=17cm]{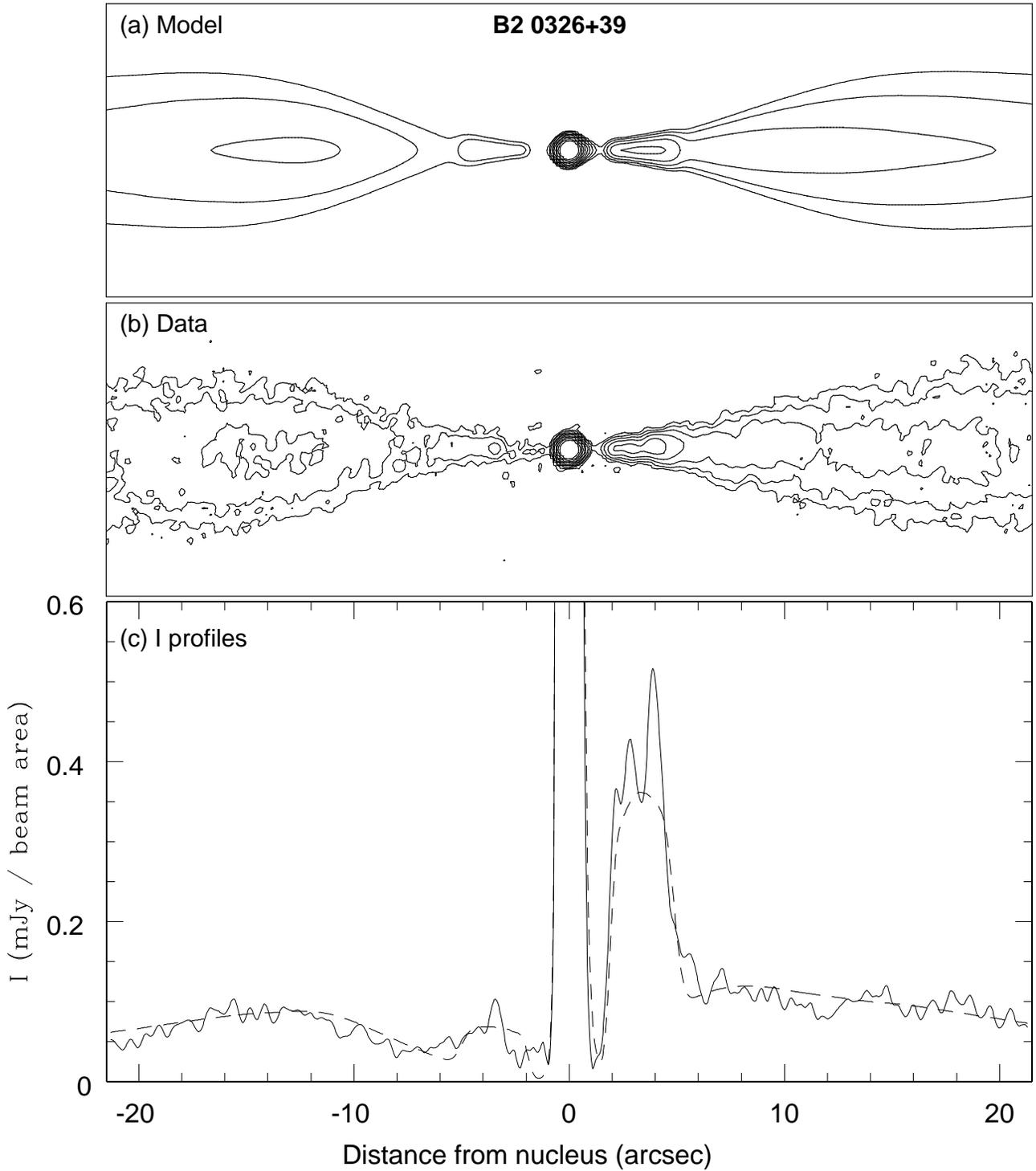}
\caption{A comparison of the model and data for B2\,0326+39  in total
intensity at 0.50 arcsec resolution for the inner 21 arcsec of each
jet. (a) model contours; (b) observed contours. The contour levels in
panels (a) and (b) are 1, 2, 4, 8, 16, 32, 64, 128, 256, 512$\times$
20\,$\mu$Jy/beam area. (c) Profiles along the jet axis for the 
data (solid line) and model (dashed line).
\label{fig:0326li}}
\end{figure*}

\subsection{Uniqueness and errors in parameter estimation}

The uniqueness of such a complex model is a concern, especially given
that the downhill simplex algorithm is not guaranteed to converge to a
global $\chi^2$ minimum. In addition, the parameters are clearly not
independent, and coupling between them leads to some indeterminacy. We
are confident of the reliability of our models, however, for the
following reasons:
\begin{enumerate}
\item We fit to deep, well-resolved images in $I$, $Q$ and $U$. These
provide a large number of independent data points with good
signal-to-noise ratio.
\item We have experimented with a wide range of starting conditions
  and have been unable to find satisfactory solutions which differ
  significantly from those presented here: models even remotely
  resembling the data are hard to find.
\item 
Whilst changes in some parameters can, in part, be offset by the variation
 of others, the extent to which this can occur is strictly limited by the
 fact that most parameters only affect the fit in limited regions and are
 well constrained by gross features in the brightness distributions.
\end{enumerate}
We derive rough uncertainties, as in LB, by varying individual
parameters until the increase in $\chi^2$ for total intensity or
linear polarization in either the flaring or the outer region
corresponds to the formal 99\% confidence level for independent
Gaussian errors. These estimates are crude (they neglect coupling
between parameters), but in practice give a good representation of the
range of qualitatively reasonable models.

\section{Comparison between models and data}
\label{results}

\begin{figure}
\includegraphics[width=8.3cm]{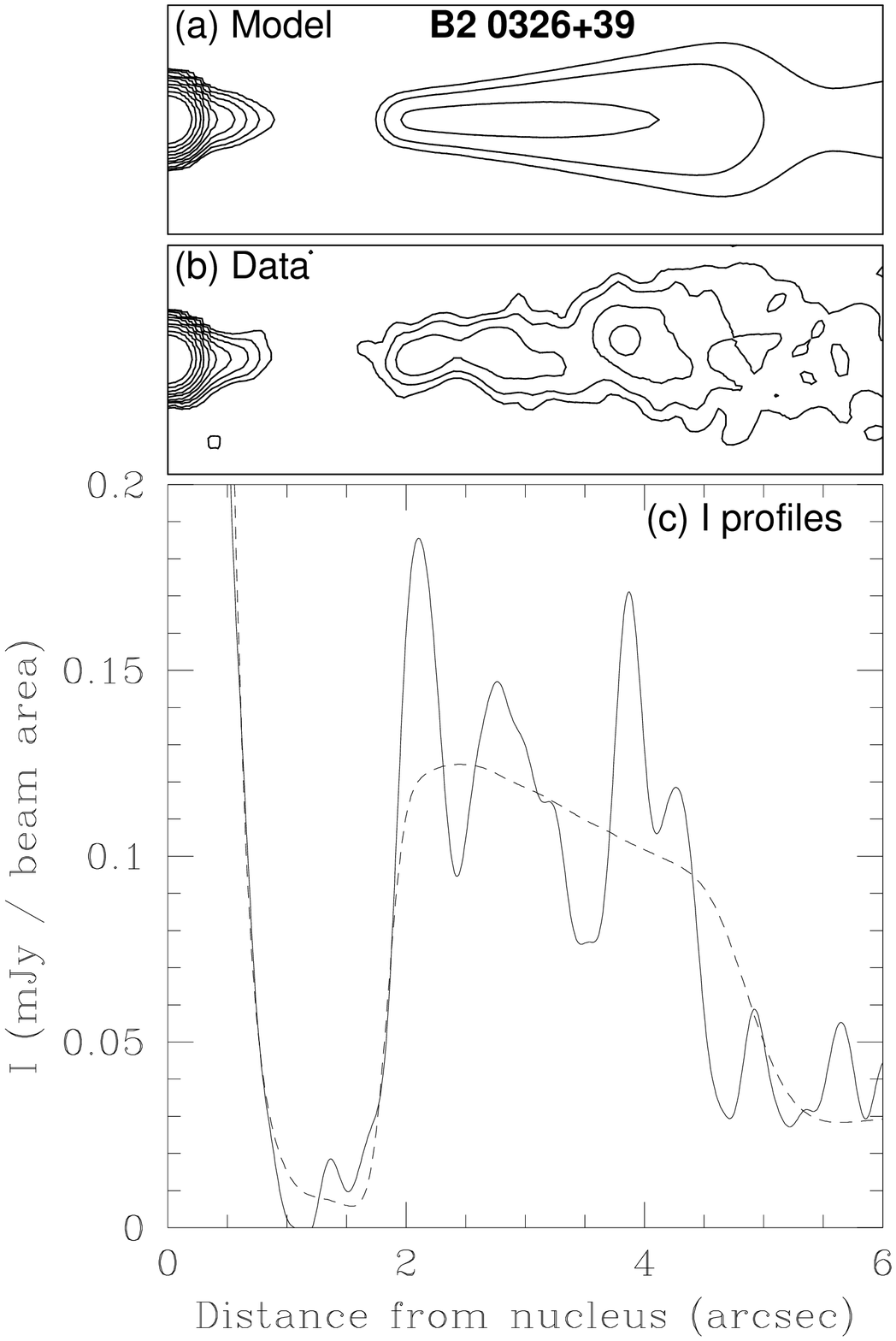}
\caption{A comparison of the model and data in total
intensity at 0.25 arcsec resolution for the inner 6 arcsec of the main
jet in B2\,0326+39. (a) model contours; (b) observed contours.
The levels in panels (a) and (b) are 1, 2, 4, 8, 16, 32, 64, 128, 256, 512$\times$
25\,$\mu$Jy/beam area. (c) Profiles along the jet ridge-line for 
the data (solid line) and model (dashed line).
\label{fig:0326hi}}
\end{figure}

\begin{figure}
\includegraphics[width=8.3cm]{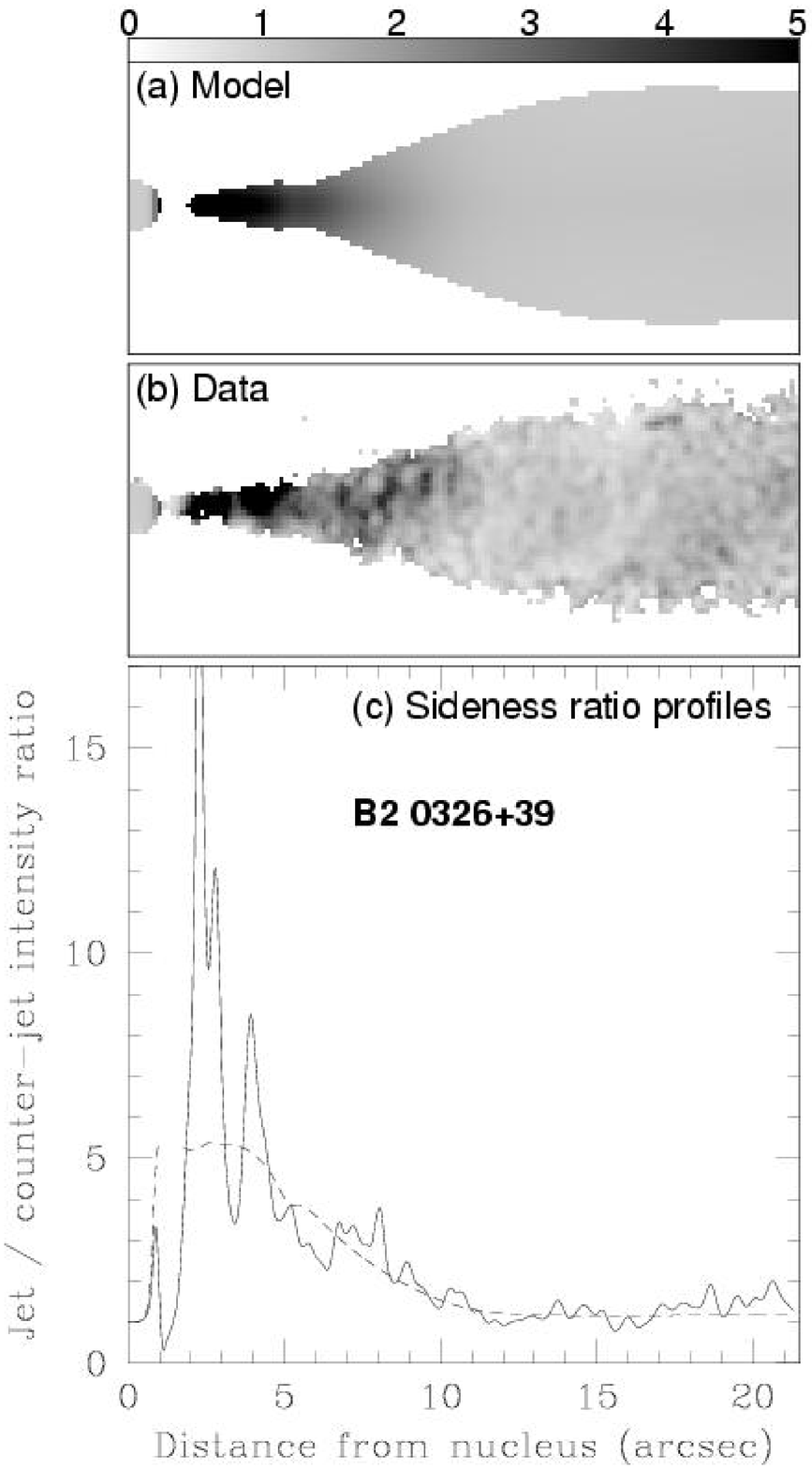}
\caption{A comparison of the model and observed jet/counter-jet sidedness
ratios at 0.50 arcsec for B2\,0326+29. (a) Model grey-scale; (b)
observed grey-scale. The grey-scale range for panels (a) and (b) is 0
-- 5, as indicated by the labelled bar. (c) Profiles along the jet
ridge-line for the data (solid line)
and model (dashed line).
\label{fig:0326ls}}
\end{figure}

\subsection{$\chi^2$ values}

The values of reduced $\chi^2$ for the final optimized models of the
two sources are listed in Table~\ref{tab:chisq}. Given that the
``noise levels'' are probably underestimated (Section~\ref{Fit-opt}),
these represent very good fits. In the remainder of this section, we
compare the models and data in detail.

%----------------------- ****** 0326 RESULTS ****** -------------------------%
\subsection{B2\,0326+39}

\subsubsection{Total intensity images and profiles}

Figs \ref{fig:0326li} and \ref{fig:0326hi} show model and observed
contour plots and profiles along the jet axis for Stokes $I$ at 0.50
and 0.25\,arcsec resolution, respectively. The model and observed
sidedness ratios are shown as grey-scale plots and on-axis profiles
in Fig.~\ref{fig:0326ls}.

\subsubsection{Fitted total intensity features}

Our optimized model successfully reproduces the following
features of the total-intensity distribution in B2\,0326+39,
as illustrated in Figs~\ref{fig:0326li} -- \ref{fig:0326ls}:
\begin{enumerate}
\item The base of the main jet is initially bright but fades rapidly
away from the nucleus and becomes invisible after 1\,arcsec.
\item Apart from the bright base of the main jet, both jets are very
faint within 2\,arcsec of the nucleus.
\item The main jet has a very bright section between 2 and 4.5\,arcsec
from the nucleus, brightening and fading very quickly at each end.
\item A similar, but much fainter, peak in the brightness of the
  counter-jet is also reproduced.
\item The main jet beyond 5\,arcsec has a very flat profile, in
contrast to the counter-jet which peaks at $\approx$15\,arcsec from the nucleus.
\item The jets are initially well collimated, flare 5 arcsec from the
nucleus and recollimate at $\approx$15 arcsec.
\item The on-axis sidedness ratio falls over the first half of the
modelled region to near unity at 11\,arcsec. This value is maintained
over the outer 10\,arcsec.
\end{enumerate}

\subsubsection{Polarization images and profiles}
\label{P-comparison}
\begin{figure*}
\includegraphics[width=15cm]{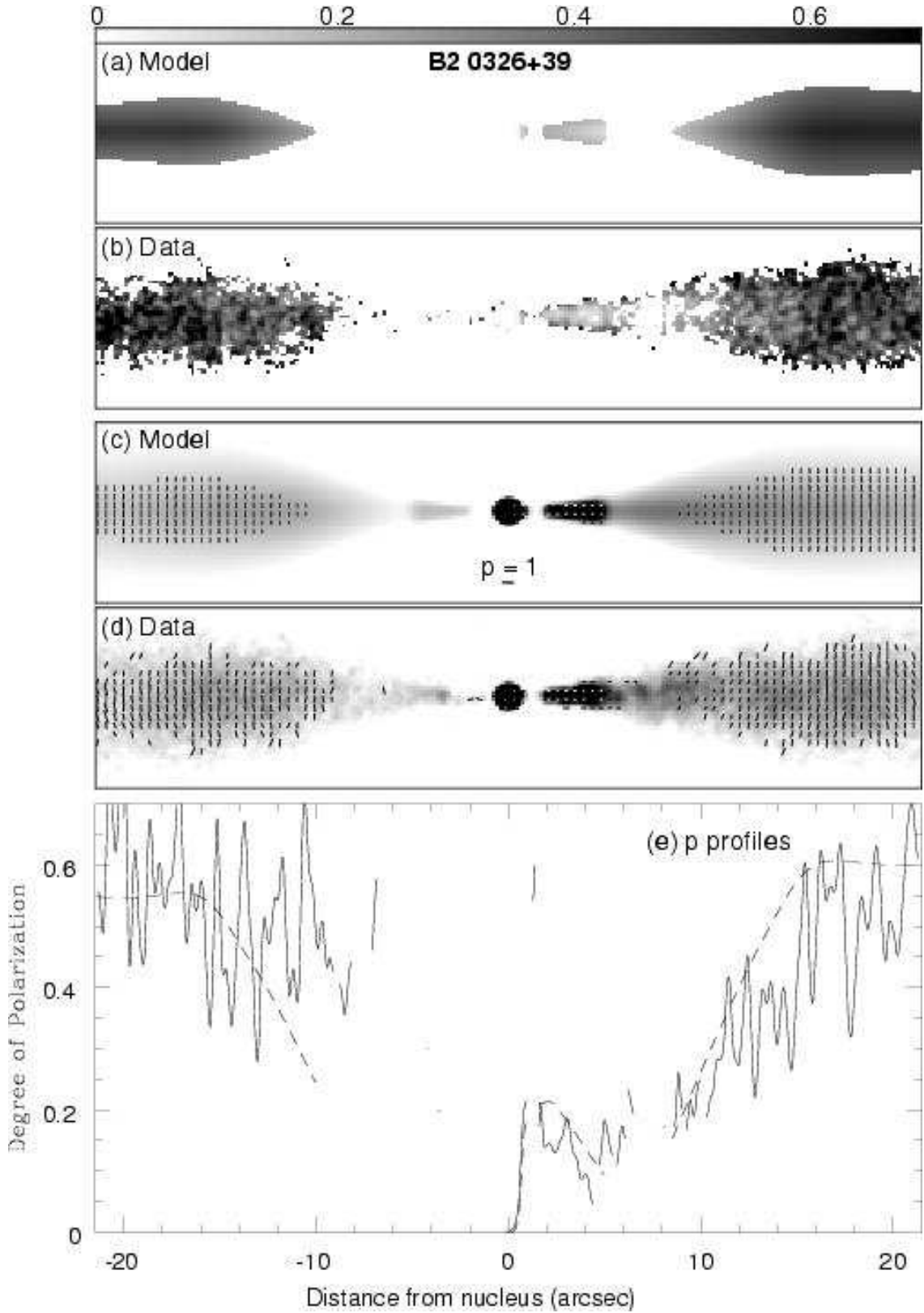}
\caption{A comparison of the model and observed polarization
  distributions at 0.50 arcsec resolution in the inner $\pm$21 arcsec
  of the jets in B2\,0326+39.  (a) and (b): grey-scales of the degree
  of polarization, $p$. The grey level is indicated by the labelled
  bar at the top of the figure. (a) model; (b) data. (c) and (d):
  vector plots showing the degree and direction of linear polarization
  superimposed on grey-scales of total intensity. The lengths of the
  vectors are proportional to $p$ (on a scale given by the labelled
  bar in panel c) and their directions are those of the apparent
  magnetic field. (c) model; (d) data. (e) Profiles of degree of
  polarization along the centre-line of the jet for the data (solid
  line) and model (dashed line). In all displays, the data and models
  have been blanked wherever the polarized power is $<3\sigma_P$
  or the total intensity is $<5\sigma_I$, using the values of
  off-source noise given in Table~\ref{noise}.
\label{fig:0326lp}}
\end{figure*}

\begin{figure}
\includegraphics[width=8.3cm]{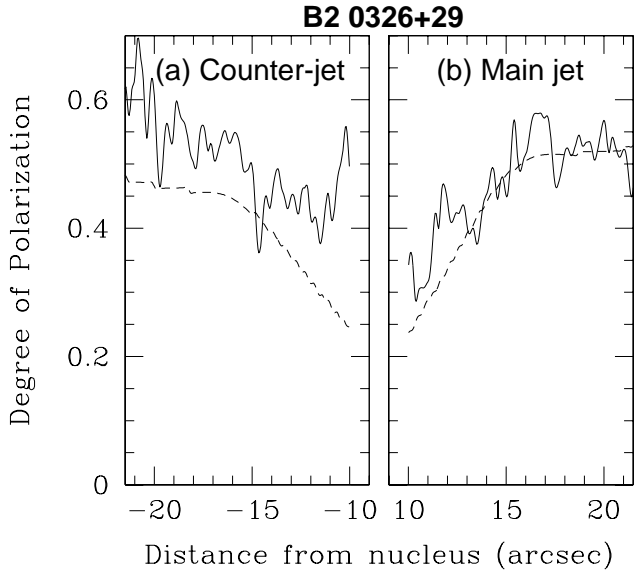}
\caption{Profiles of the degree of polarization averaged across the
jets of B2\,0326+39. Full line:data; dashed line: model. The plots
cover the range 10 -- 21.5\,arcsec from the nucleus, where there is
little variation across the jets (Fig.~\ref{fig:0326tp}). (a)
counter-jet; (b) main jet. The resolution is 0.5\,arcsec
FWHM. \label{fig:0326avlp}}
\end{figure}

\begin{figure}
\includegraphics[width=8.3cm]{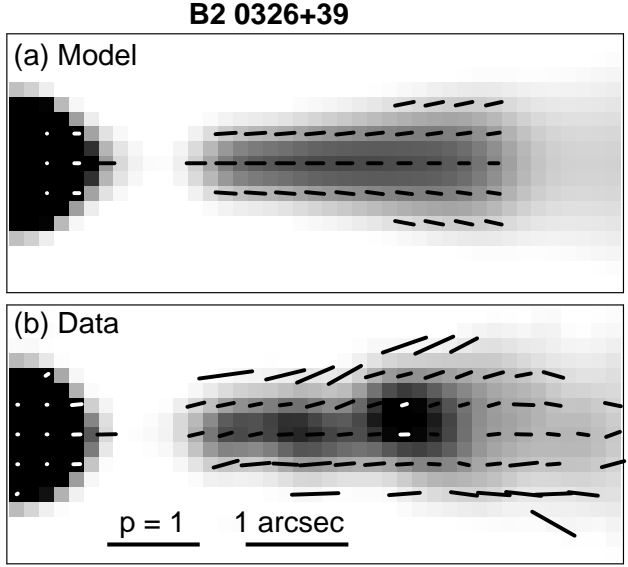}
\caption{Vector plots of the degree and direction of linear
  polarization superimposed on grey-scales of total intensity for the
  inner 6\,arcsec of the main jet in B2\,0326+39 at a resolution of
  0.50\,arcsec. The vector lengths are proportional to the degree of
  polarization, $p$ and their directions are those of the apparent
  magnetic field. (a) model; (b) data.\label{fig:0326lpsub}}
\end{figure}

Fig.~\ref{fig:0326lp} illustrates the observed and model polarization
structures of the jets in B2\,0326+39 as grey-scales of the degree of
polarization, $p$, vector plots representing $p$ and the apparent magnetic
field direction and profiles of $p$ along the jet centre-line. There is
little transverse variation of the degree or direction of polarization
across the jets at large $z$ 
(Figs~\ref{fig:0326lp}b and d) and the ridge-line profiles are
quite noisy, so we also show profiles of $p$ averaged across the jets from
10 -- 21.5\,arcsec from the nucleus in Fig.~\ref{fig:0326avlp}.  Vector
plots for the inner regions of the main jet are shown on an expanded scale
in Fig.~\ref{fig:0326lpsub}. The 0.25\,arcsec resolution images have low
polarized signal-to-noise ratio and are not shown.

\begin{figure}
\includegraphics[width=8.3cm]{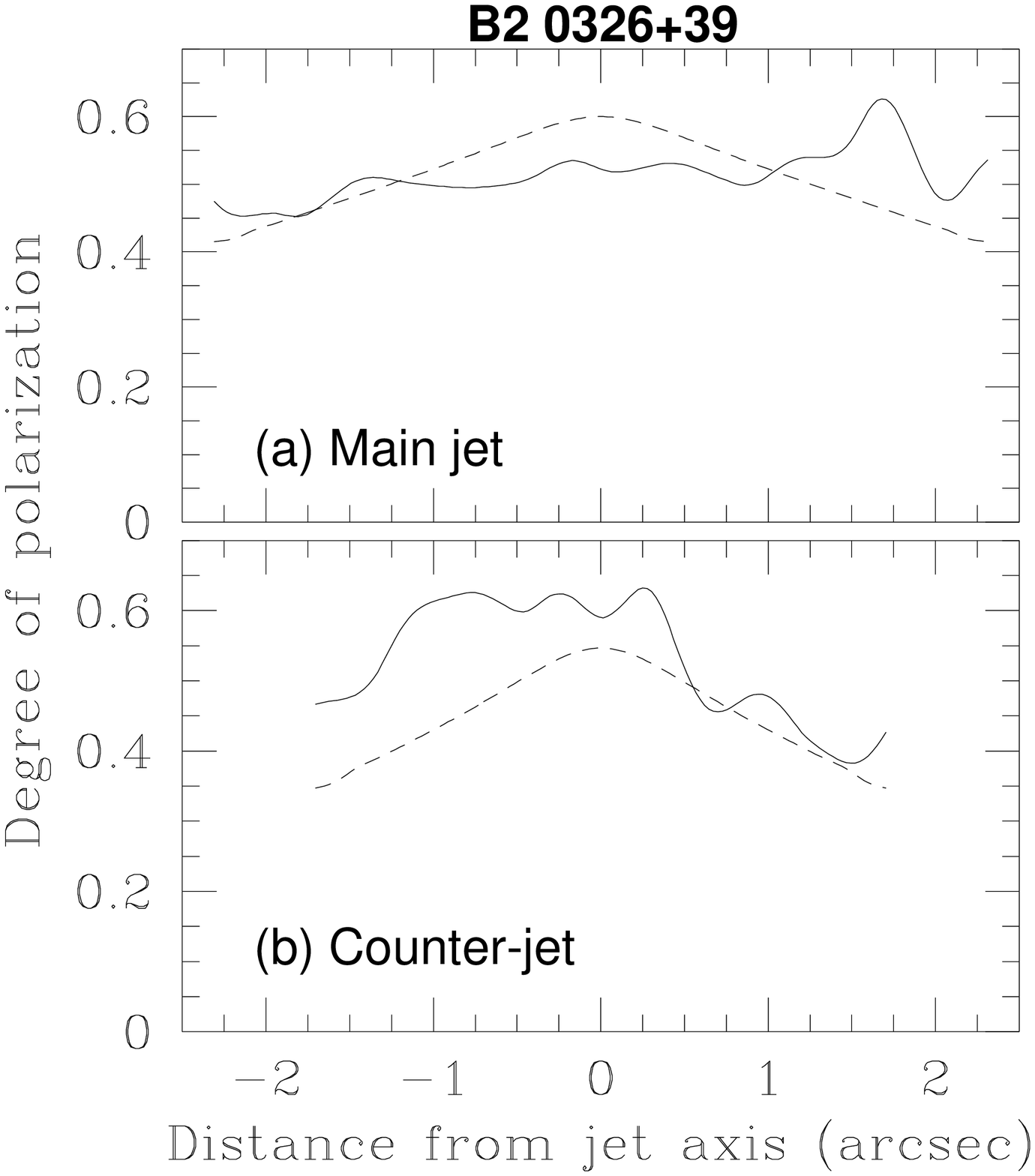}
\caption{Transverse profiles of the degree of polarization in
B2\,0326+39 obtained by averaging over $z$ 
between 15 and 21\,arcsec. Full line: data; dashed line: model.  The
observed profiles are derived by averaging over images of $p$ made
without blanking, in order to avoid biasing the results to high
values.  The resolution is 0.5\,arcsec FWHM.
\label{fig:0326tp}}
\end{figure}

Individual transverse profiles of $p$ are also noisy, so Fig.~\ref{fig:0326tp}
shows averages over the outer parts of the main and
counter-jets (between 15 and 21\,arcsec from the nucleus), where there
is little longitudinal variation.

\subsubsection{Fitted polarization features}

Our model is successfully able to reproduce the following features in
the polarization images (Figs~\ref{fig:0326lp}--\ref{fig:0326tp}):

\begin{enumerate}
\item The inner knot and the bright regions of the main jet within
5\,arcsec of the nucleus have longitudinal apparent magnetic field.
\item The bright part of the main jet is more polarized at its edges, 
forming a V-shaped structure in $p$ (Figs~\ref{fig:0326lp}a and b).
\item In this region,the on-axis degree of polarization falls from 0.2 at the
  base of the main jet to 0.1 at 4\,arcsec (Fig.~\ref{fig:0326lp}c).
\item The main jet has a region where no polarized signal is detected
at the 3$\sigma$ level between 5 and 9 arcsec, setting a limit of $p <
0.15$. The model distribution is consistent with this limit. In the
counter-jet, no significant polarized signal is seen closer than
10\,arcsec from the nucleus and none is predicted (the observational
limits are not severe because of the low total intensity).
\item Beyond 10\,arcsec from the nucleus, the jet
is polarized such that the apparent magnetic field vector is
transverse to the jet axis and no parallel-field edge is seen.
\item Between 10 and 15 arcsec, the polarizations of both the jet and
counter-jet increase monotonically before becoming constant with $p
\approx 0.5$ further out. [The apparent underestimation of $p$ in the
counter-jet at $\approx$10\,arcsec from the nucleus is in a region of low
polarized signal and is probably not significant].
\end{enumerate}

\begin{figure*}
\includegraphics[width=17cm]{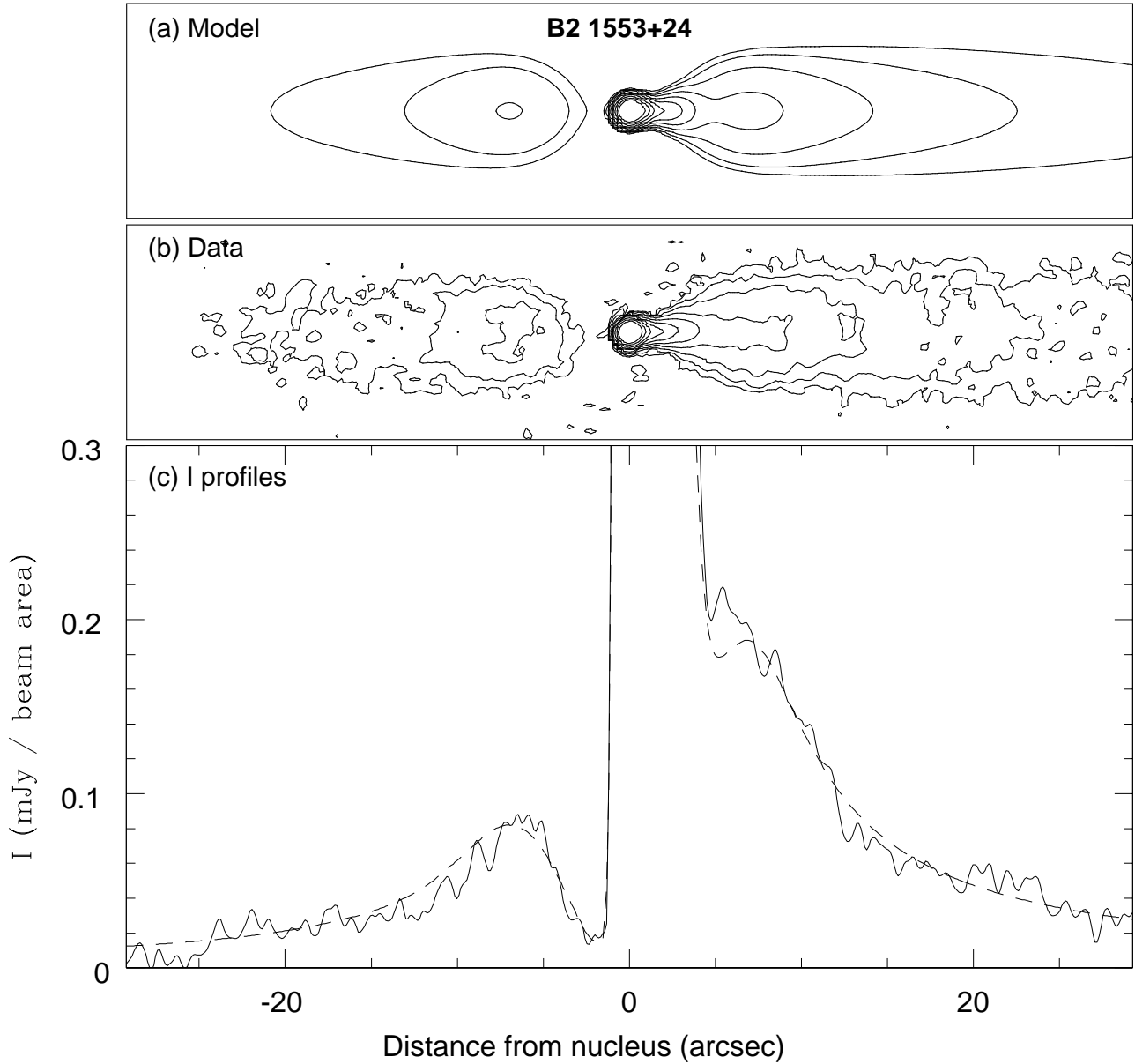}
\caption{A comparison of the model and observed total intensity for
B2\,1553+24 at 0.75 arcsec resolution. The plots show the inner
$\pm$29\,arcsec.  (a) and (b): contour plots of total intensity with
levels at 1, 2, 4, 8, 16, 32, 64, 128, 256, 512$\times$
20\,$\mu$Jy/beam area. (a) model; (b) data. (c) Profiles along the jet
centre-line for the data (solid line) and model (dashed line).
\label{fig:1553li}}
\end{figure*}

\subsubsection{Features that cannot be fitted well}

Our model cannot reproduce small-scale and/or non-axisymmetric
features, the most prominent of which are the three knots of emission
in the bright region of the main jet between 2 and 4.5\,arcsec from
the nucleus (Fig~\ref{fig:0326hi}). The model correctly averages the
brightness in this region into a smooth profile. In addition, there
are a number of small, but significant discrepancies between model and
data:
\begin{enumerate}
\item The inner region of the counter-jet is observed to be brighter
than that of the main jet (Fig~\ref{fig:0326li}b), an effect that is
impossible to reproduce using intrinsically identical jets. This may
also be caused by fluctuations on a small scale (e.g. a single knot in
the counter-jet).
\item The counter-jet first brightens significantly 1\,arcsec further from
the nucleus than the main jet and the minimum in the total intensity
profile at the end of the flaring region is 2\,arcsec further from the
nucleus in the counter-jet (Fig~\ref{fig:0326li}c). This is not
reproducible by our model, for which the overall shape of the flaring
region should be similar for the two jets, even if their brightnesses
differ significantly.
\item The counter-jet intensity decreases with
$z$ faster than the model predicts 
for $z > $15\,arcsec. This has the
effect of increasing the sidedness ratio slightly in this region, an
effect which is inconsistent with a monotonically decelerating flow
(see Section \ref{acceleration}).
\item The model predicts a degree of polarization which is slightly higher
for the main jet than the counter-jet at large $z$. The converse is true,
although the discrepancy is marginal (Figs~\ref{fig:0326lp}e and
\ref{fig:0326avlp}).
\item In the outer parts of the modelled region, $p$ is predicted to
  be higher on-axis than at the jet edges. This is marginally
  inconsistent with the data for the main jet, where $p$ has little if
  any transverse variation (Fig.~\ref{fig:0326tp}).
\end{enumerate}

%------------------------ ****** 1553 RESULTS ****** ---------------------------%
\subsection{B2\,1553+24}

\subsubsection{Total intensity images and profiles}

\begin{figure}
\includegraphics[width=8.3cm]{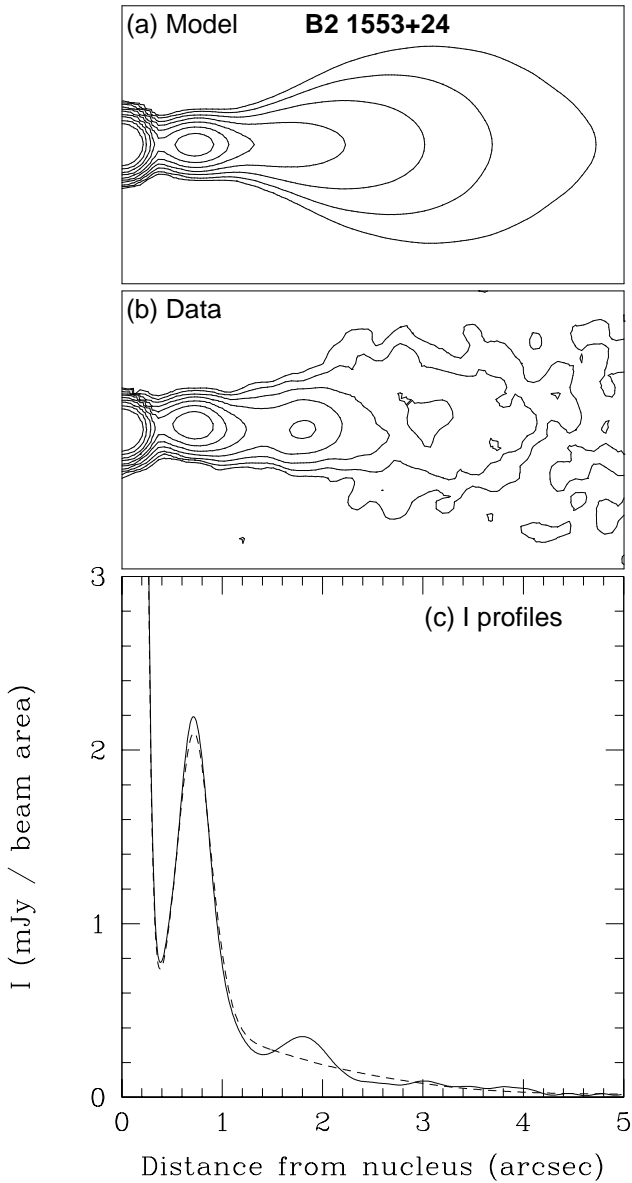}
\caption{A comparison of the model and observed total intensity at
  0.25\,arcsec resolution for the inner 5\,arcsec of the main jet in
  B2\,1553+24. (a) and (b): contour plots with levels at 1, 2, 4, 8,
  16, 32, 64, 128, 256, 512$\times$ 20\,$\mu$Jy/beam area. (a) model; (b) data.
 (c) A  profile along the jet centre-line showing the observed (solid line) and
  model (dashed line) values.
\label{fig:1553hi}}
\end{figure}

\begin{figure}
\includegraphics[width=8.3cm]{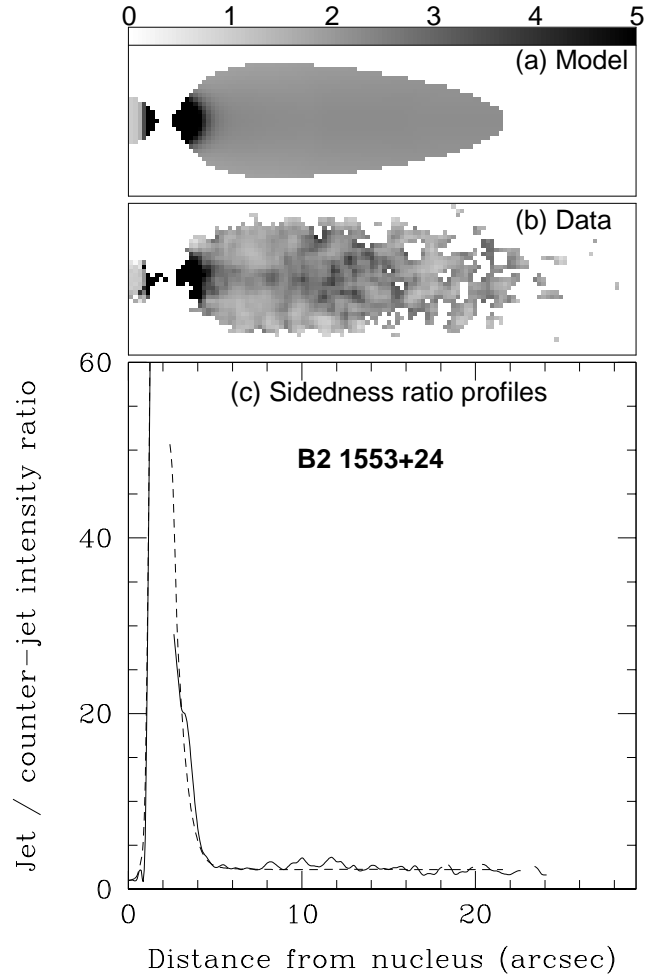}
\caption{A comparison of the model and observed jet/counter-jet
sidedness ratios at 0.75 arcsec for B2\,1553+24. (a) and (b):
grey-scale images of sidedness ratio, in the range 0 -- 5 as indicated
by the labelled bar. (a) model; (b) data.(c) sidedness profiles along
the centre-line of the jet showing data (full line) and model (dashed
line).
\label{fig:1553ls}}
\end{figure}

Figs \ref{fig:1553li} shows contours and longitudinal profiles of
total intensity for the model and observed images at 0.75 arcsec
resolution. Corresponding plots for the inner 5\,arcsec of the main
jet at 0.25-arcsec resolution are shown in Fig.~\ref{fig:1553hi} (the
counter-jet is invisible close to the nucleus on these images).  The
model and observed sidedness ratios are compared in
Fig.~\ref{fig:1553ls}.

\subsubsection{Fitted total intensity features}

The following features of the jets in B2\,1553+24 are fitted accurately by
our optimized model:

\begin{enumerate}
\item The main jet initially brightens very rapidly away from the 
 nucleus, peaking at 0.7\,arcsec separation and then fading to give
 the appearance of a bright knot.
\item This jet remains well-collimated for 1.5\,arcsec before flaring
abruptly.
\item At $z \approx$ 7\,arcsec, both jets recollimate, maintaining an
approximately constant width at large $z$.
\item The counter-jet is not seen in the inner 3\,arcsec of the 0.75-arcsec
resolution image. After this point it brightens, reaching a peak $\approx$7
arcsec from the nucleus.
\item At $z \approx$ 6\,arcsec in the main jet, the total
intensity profile flattens before falling again from $z \approx$ 8\,arcsec.
\item The jet/counter-jet sidedness ratio drops rapidly with $z$, reaching a
constant value of $\approx$2.2 at $z \approx $ 6\,arcsec.
\end{enumerate}

\subsubsection{Polarization images and profiles}

\begin{figure*}
\includegraphics[width=17cm]{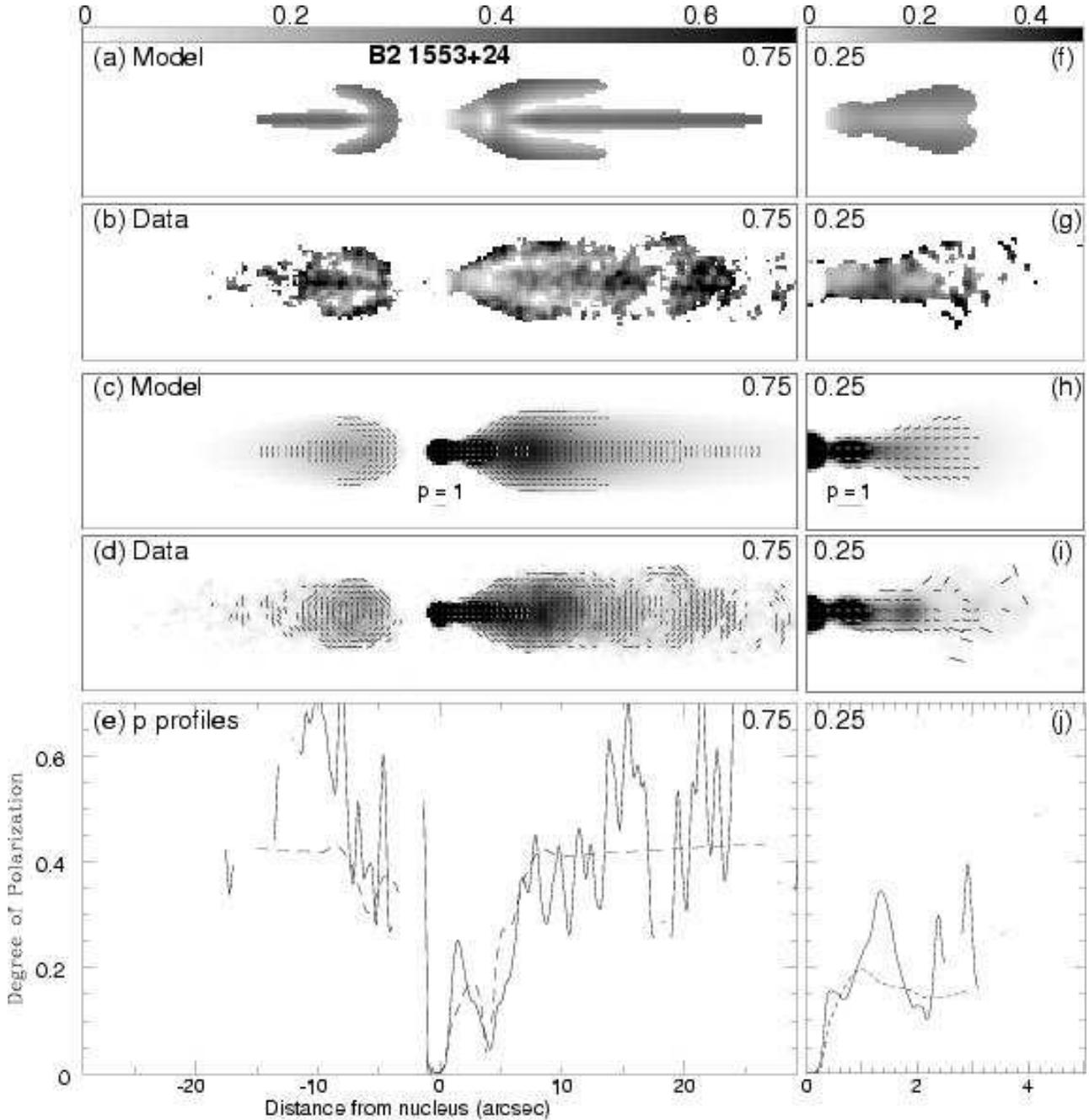}
\caption{A comparison of the model and observed polarization for
B2\,1553+24. Panels (a) -- (e) show the inner $\pm$29\,arcsec at
0.75\,arcsec resolution. (a) and (b): grey-scale images of the degree
of polarization, $p$, in the range 0 -- 0.7, as indicated by the
labelled wedge. (a) model; (b) data. Panels (c) and (d) are vector
plots showing the degree of polarization and the apparent magnetic
field direction. The vector lengths are proportional to $p$, on a
scale indicated by the bar on panel (c). (c) model; (d) data. (e)
Profiles of the degree of polarization along the centre-line of the
jet for model (dashed) and data (full). The data and models have been
blanked wherever the polarized signal is $<3\sigma_P$ or the total
intensity is $<5\sigma_I$, using the values of off-source noise as
given in Table~\ref{noise}. Panels (f) -- (j): as (a) -- (e), but for
the inner 5\,arcsec of the main jet at a resolution of
0.25\,arcsec. Note that the grey-scale range in panels (f) and (g) is
0 -- 0.5 and therefore differs from that in panels (a) and (b).
\label{fig:1553lp}}
\end{figure*}

\begin{figure}
\includegraphics[width=8.3cm]{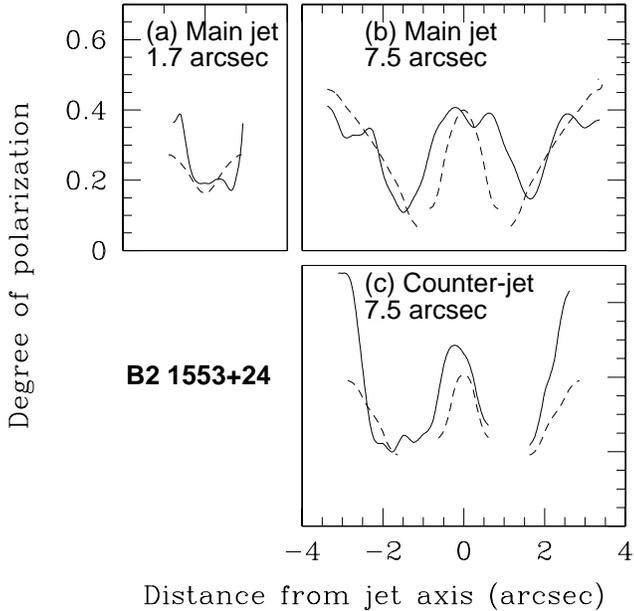}
\caption{Transverse profiles of the degree of polarization in
B2\,1553+24. (a) Profile at a distance of 1.7\,arcsec from the nucleus
in the main jet. The apparent field is parallel to the jet axis across
the entire width. (b) and (c) Profiles at 7.5\,arcsec from the nucleus
for the main and counter-jets, respectively. In both jets, the
apparent field is parallel to the jet axis near the jet edge and
perpendicular towards the centre. Full line: data; dashed line:
model. The profiles are blanked as in Fig.~\ref{fig:1553lp} and the
resolution is 0.75\,arcsec.
\label{fig:1553tp}}
\end{figure}

Fig \ref{fig:1553lp} shows grey-scale images of the degree of
polarization $p$ and vector images showing the orientation of the
apparent magnetic field at both 0.75 and 0.25 arcsec resolution,
together with profiles of $p$ along the jet axis. Transverse profiles
of the degree of polarization at two representative points in the jet
and one in the counter-jet are shown in Fig \ref{fig:1553tp}.

\subsubsection{Fitted polarization features}

Our model is successfully able to reproduce the following features
of the linear polarization in B2\,1553+24:
\begin{enumerate}
\item Within 3.5\,arcsec of the nucleus, the degree of polarization is
  approximately constant on-axis, with $p \approx 0.15$.  It increases
  with distance from the axis, reaching $\approx$0.3 at the edge of
  the jet.
\item In this region the apparent field is orientated approximately
parallel to the jet axis.
\item At $\approx$4 arcsec from the nucleus the polarization
on-axis in the main jet is very low.
\item Between 4 and 8 arcsec from the nucleus the on-axis polarization
in the main jet increases to $p \approx$0.4; further out it remains
approximately constant at that value.
\item The counter-jet polarization on-axis remains approximately
constant where it can be measured (4 - 15 arcsec)
\item Further than 4\,arcsec from the nucleus, both the jet and
counter-jet are polarized with a transverse apparent field  on-axis.
\item The apparent field is longitudinal along the edges of both jets
  wherever there is significant signal.
\item There is a semicircular arc of enhanced polarized emission with
  a circumferential apparent field in the counter-jet, between 4 and 8
  arcsec from the nucleus.
\end{enumerate}

\subsubsection{Features that cannot be fitted well}

Our model cannot fit the following features:
\begin{enumerate}
\item The bright knot at the base of the main jet does not lie exactly
on the jet axis and cannot be fitted precisely by our axisymmetric model.
\item There is a second knot 1.9\,arcsec from the nucleus, where the
jet starts to expand rapidly (Fig.~\ref{fig:1553hi}b and c). It would
be possible to fit this feature by adjusting the emissivity profile to
be much flatter locally, in which case the increase in path length
through the jet would naturally generate a maximum. In the absence of
any constraint from the counter-jet, this complication cannot be justified.
\item The arc of emission $\approx$20 arcsec from the nucleus in the main
jet, particularly noticeable in the polarized images, and similar to a
prominent feature seen in 3C31 (LB), cannot be fitted.
\end{enumerate}

\section{Physical parameters}
\label{physical}

%--------------------- ****** MODEL INTERPRETATION ****** ------------------------%
\subsection{Model parameters and confidence limits}

The following section examines the structures of the jets implied by our models. 
All  distances from the nucleus are given
in linear units in the jet frame (i.e.\ {\em not} in projection on the sky).

Table~\ref{Params} gives the fitted parameters and error estimates,
defined as in  Section \ref{model}, for both sources.

\begin{table*}
\caption{Fitted parameters and error estimates.\label{Params}}
\begin{minipage}{120mm}
\begin{tabular}{llrrrcrrr}
\hline
&&\multicolumn{3}{c}{0326+39}&&\multicolumn{3}{c}{1553+24} \\
Quantity                 &     Symbol     &~opt~ &~min\footnote{The Symbol $<$ means that any value smaller than the quoted maximum is allowed.}  &~max\footnote{The Symbol $>$ means that any value larger than the quoted minimum is allowed.}  &&~opt~ &~min  &~max  \\
\hline
Angle to line of sight (degrees)& $\theta$& 63.8 & 59.0 & 68.9 &&  7.7 &  6.5 &  9.0 \\
&&&&&&&&\\
Geometry                                  &&&&&&&&\\
~~Boundary position (kpc)       &  $r_0$  &10.02 & 9.52 &10.60 &&48.15 &44.56 &51.57 \\
~~Jet half-opening angle (degrees)&$\xi_0$         
                                          & 2.37 &~$<$~~& 6.75 && 0.75 & 0.53 & 1.32 \\
~~Width of jet at outer boundary (kpc)&$x_0$& 2.47 & 2.28 & 2.64 && 3.36 & 3.09 & 3.75 \\
&&&&&&&&\\
Velocity                                  &&&&&&&&\\
~~Boundary positions (kpc)                &&&&&&&&\\
~~~~inner                &$\rho_{\rm v_1}$& 2.43 & 0.68 & 4.10 &&21.01 &10.78 &25.71 \\
~~~~outer                &$\rho_{\rm v_0}$& 6.17 & 4.88 & 7.53 &&28.04 &17.69 &36.38 \\
~~On~$-$~~axis velocities / $c$           &&&&&&&&\\
~~~~inner                &   $\beta_1$    & 0.80 & 0.64 & 0.93 && 0.74
& 0.70 &~$>$\footnote{Undetermined, as the counter-jet is not  visible in this region.}~    \\
~~~~outer                &   $\beta_0$    & 0.09 &~$<$~~& 0.18 && 0.17 & 0.13 & 0.21 \\
~~Fractional velocity at edge of jet      &&&&&&&&\\
~~~~inner                &     $v_1$      & 0.60 & 0.28 & 1.01 && 0.83 & 0.68 & 0.95 \\
~~~~outer                &     $v_0$      & 0.65 &~$<$~~& 4.34 && 0.63 &
0.23 &~$>^{\rm c}$\\
&&&&&&&&\\
Emissivity                                &&&&&&&&\\
~~Boundary positions (kpc)                &&&&&&&&\\
~~~~inner                &$\rho_{\rm e_1}$& 1.05 & 0.73 & 1.19 && 0.00 &      &      \\
~~~~2                    &$\rho_{\rm e_2}$& 2.68 & 2.62 & 2.77 && 4.87 & 4.66 & 5.09 \\
~~~~3                    &$\rho_{\rm e_3}$& 3.00 & 2.93 & 3.05 && 7.58 & 7.23 & 7.89 \\
~~~~4                    &$\rho_{\rm e_4}$&\multicolumn{3}{c}{not used}&&23.13 &20.27 &26.02 \\
~~On~$-$~~axis emissivity exponents       &&&&&&&&\\
~~~~inner                &     $E_1$      & 6.08 & 5.03 & 6.41 &&      &      &      \\
~~~~2                    &     $E_2$      & 2.39 & 1.52 & 2.83 && 0.00 & 0.00 & 0.88 \\
~~~~3                    &     $E_3$      &12.15 &10.66 &15.09 && 8.38 & 7.81 & 8.85 \\
~~~~4                    &     $E_4$      & 1.27 & 1.08 & 1.43 && 3.34 & 3.15 & 3.52 \\
~~~~5                    &     $E_5$      &      &      &      && 1.87 & 1.72 & 2.04 \\
~~Fractional emissivity at edge of jet    &&&&&&&&\\
~~~~inner boundary       &     $e_1$      & 0.61 & 0.07 & 1.29 && 0.61 &~$<$~~& 2.23 \\
~~~~boundary 2           &     $e_0$      & 0.24 & 0.12 & 0.37 && 0.25 & 0.14 & 0.46 \\
~~Emissivity ratio at inner boundary& $g$ & 0.022& 0.005& 0.036&&\multicolumn{3}{c}{not used}\\
&&&&&&&&\\
B-field                                   &&&&&&&&\\
~~Boundary positions (kpc)                &&&&&&&&\\
~~~~inner                &$\rho_{\rm B_1}$& 0.92 &~$<$~~& 2.95 &&30.37 &18.63 &37.46 \\
~~~~outer                &$\rho_{\rm B_0}$& 8.66 & 7.20 & 9.99 &&38.50 &32.55 &44.13 \\
~~RMS field ratios                        &&&&&&&&\\
~~~~radial/toroidal                       &&&&&&&&\\
~~~~~~inner region       &     $j_1$      & 0.81 & 0.27 & 1.38 && 0.68 &~$<$~~& 1.07 \\
~~~~~~outer region       &     $j_0$      & 0.53 & 0.30 & 0.72 && 0.29 &~$<$~~& 0.55 \\
~~~~longitudinal/toroidal                 &&&&&&&&\\
~~~~~~inner region       &     $k_1$      & 1.43 & 1.18 & 1.74 && 3.06 & 2.60 & 3.62 \\
~~~~~~outer region       &     $k_0$      & 0.11 &~$<$~~& 0.26 && 0.00 &~$<$~~& 0.53 \\
\hline
\end{tabular}
\end{minipage}
\end{table*}

\subsection{B2\,0326+39}
\label{0326model}

\begin{figure}
\includegraphics[width=7.6cm]{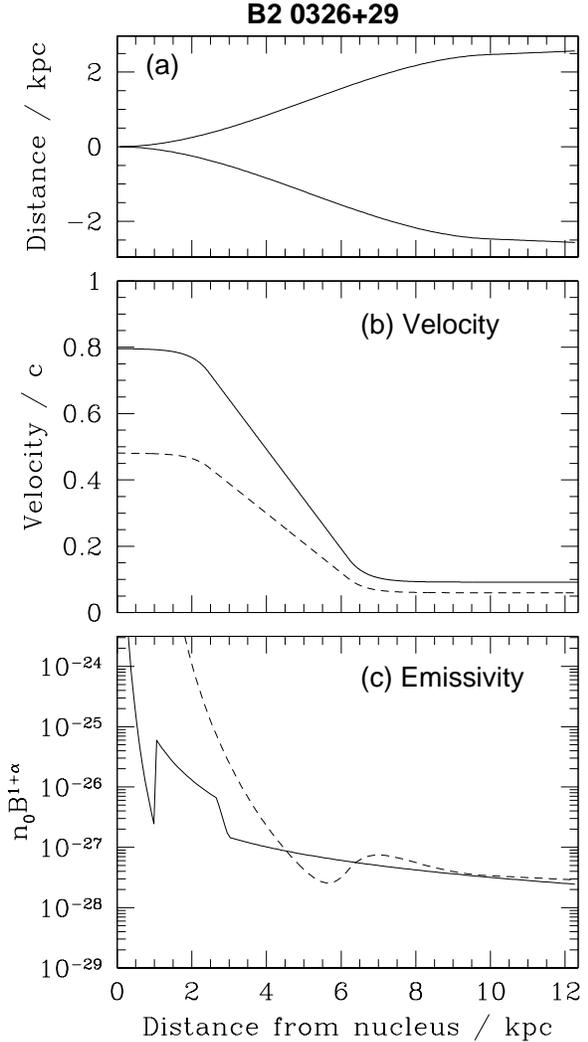}
\caption{Profiles of intrinsic parameters along the jets of B2\,0326+39 as 
functions of $z$ (not projected on the sky). (a) the
shape of the jet edge (the scales are the same for both axes). (b) the
velocity profile along the jet axis (solid line) and jet edge (dashed
line). (c) the on-axis emissivity profile, converted to $n_0 B^{1+\alpha}$
with $n_0$ in m$^{-3}$ and $B$ in T. Solid line: model; dashed line: adiabatic
approximation with the magnetic-field evolution expected from flux
freezing. The normalization for the adiabatic profile is set so that it
agrees with the free model at large distances from the nucleus.
\label{fig:0326profiles}}
\end{figure}

\subsubsection{Geometry and angle to the line of sight}
\label{0326angle}

Our best-fitting model requires an angle to the line of sight of $64^\circ
\pm 5^\circ$.  The jet is initially well collimated, then flares between 3
and 8 kpc from the nucleus, reaching a maximum opening angle of
$\approx$20$^\circ$ (Fig.~\ref{fig:0326profiles}a). Further from the
nucleus, it recollimates to a cone with an opening angle of only a few
degrees.

\subsubsection{Velocity field}
\label{0326velocity}

The velocity profile along the jet axis is shown by the solid line in
Fig.~\ref{fig:0326profiles}(b). The initial velocity of $\beta \approx
0.8$ is maintained within 2\,kpc of the nucleus. The jet then decelerates
uniformly out to $\approx$7\,kpc from the nucleus, where the velocity
reaches its asymptotic value. The data are consistent with a roughly
constant velocity of $\beta \approx 0.09$ until the end of the modelled
region. There are large fractional uncertainties, since variations in
Doppler factor are slight, and we can only set an upper limit of $\beta \la
0.15$ anywhere in this region.  We discuss the velocity variations in the
outer parts of the jet in more detail in Section~\ref{acceleration}.

The transverse velocity structure is poorly constrained everywhere. In the
high-velocity region close to the nucleus, the jet is narrow and poorly
resolved transversely. At large $z$, the jet
velocity is low, so differences in sidedness ratio between the centre and
edge of the jet, which provide the principal constraint on transverse
velocity variations, are undetectably small for our preferred inclination.

\subsubsection{Emissivity}
\label{0326emissivity}

The model on-axis profile of $n_0 B^{1+\alpha}$ for B2\,0326+39
(derived from the emissivity) is shown by the solid line in Fig
\ref{fig:0326profiles}(c). In general, the profile flattens with
increasing $z$: the fitted power-law indices are 6.1 ($<$1.0\,kpc),
2.4 (1.0 -- 2.7\,kpc) and 1.3 ($>$3.0\,kpc).  There are rapid changes
in emissivity at 1.0\,kpc and 2.8\,kpc. The former is explicitly
modelled as a discontinuity, with the emissivity changing by a factor
of $1/g \approx 40$. The latter, fit by a power-law with a very steep
index between 2.7 and 3\,kpc, is also consistent with a discontinuity.

In the outer parts of the jet, the emissivity at the jet edge is
$\approx$0.25 of its on-axis value with an error of a factor of
$\approx$2. Elsewhere, the jet is poorly resolved transverse to its axis
and the off-axis variation is poorly determined.

\subsubsection{Magnetic-field structure}
\label{0326magnetic}

\begin{figure}
\includegraphics[width=8.2cm]{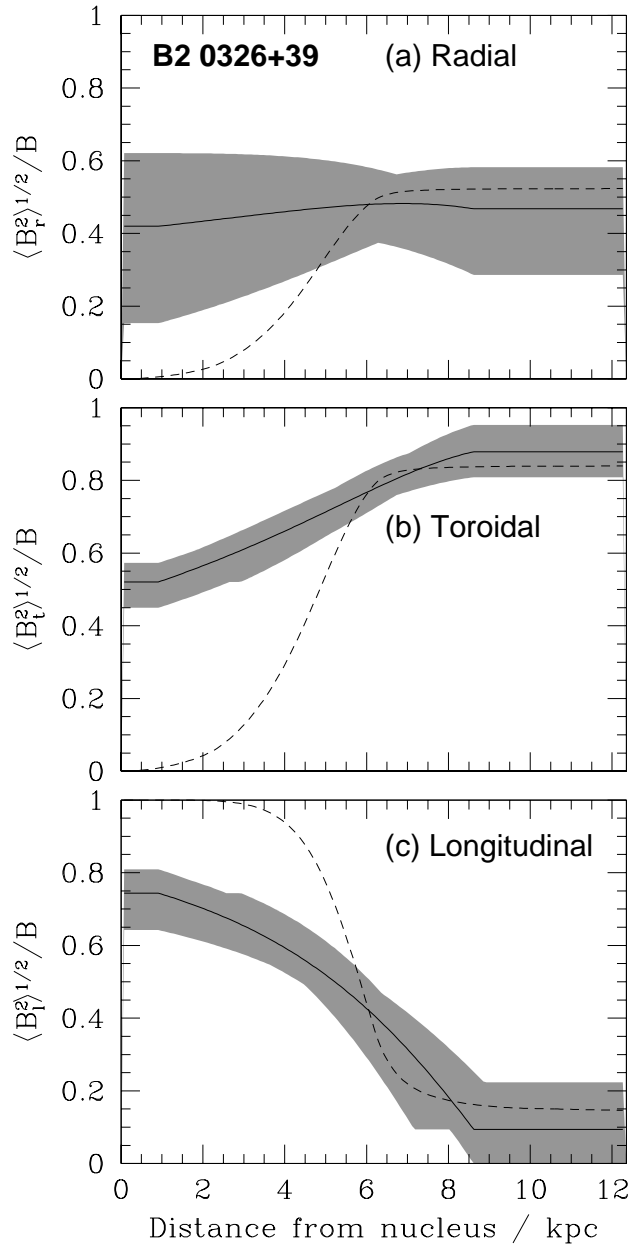}
\caption{Profiles of: (a) radial, (b) toroidal and (c) longitudinal
components of the magnetic field along the streamlines for
B2\,0326+39.  Solid line: best fit model; shaded area: error on the
best fit model derived from the limits in Table \ref{Params}; dashed line:
the profile expected on axis, in the absence of shear, if the magnetic
field is frozen into the plasma. The latter curves are normalized to match 
the model at 6\,kpc from the nucleus.
\label{fig:0326bprofiles}}
\end{figure}

Fig. \ref{fig:0326bprofiles} shows profiles of the radial, toroidal and
longitudinal components of the magnetic field.  As there is no variation
with streamline index, these curves apply to all streamlines. The solid
lines show our best-fitting model and the shaded areas the allowed
errors. Close to the nucleus, the longitudinal component dominates. As the
distance from the nucleus increases, this component drops, reaching a
value consistent with zero by 8.5\,kpc.  The largest
single component is toroidal from 3.5\,kpc outwards, the fraction of radial
field remaining roughly constant everywhere.  The field structure from
8.5\,kpc outwards essentially consists of two-dimensional field sheets
aligned perpendicular to the jet axis with a radial/toroidal field ratio
$\approx$0.6. The transverse polarization profiles for the main and
counter-jets (Fig.~\ref{fig:0326tp}) tell slightly different
stories about the details of this field configuration. In the main jet,
the profile is essentially flat, and is marginally inconsistent with the
model prediction of a higher degree of polarization on-axis. It would be
slightly better fit by a model in which the toroidal and radial components
are equal. The expected polarization for a non-relativistic jet with this
field structure at an angle $\theta$ to the line of sight is $p = 0.43$
for $\alpha = 0.55$ \citep{Laing80}. This is slightly less than the
observed value, and no transverse variation would be expected. In the
counter-jet, the degree of polarization is higher on-axis ($p \approx
0.6$) and there are hints of a decrease towards the edge. Both of these
features are better fit by increasing the toroidal component. The
optimization has compromised between the fits for the two jets to produce
the results shown in Fig.~\ref{fig:0326tp}. It is unclear whether the
differences between the field component ratios in the main and
counter-jets are real or a product of noisy data.

\subsection{B2\,1553+24}
\label{1553model}

\begin{figure}
\includegraphics[width=8.2cm]{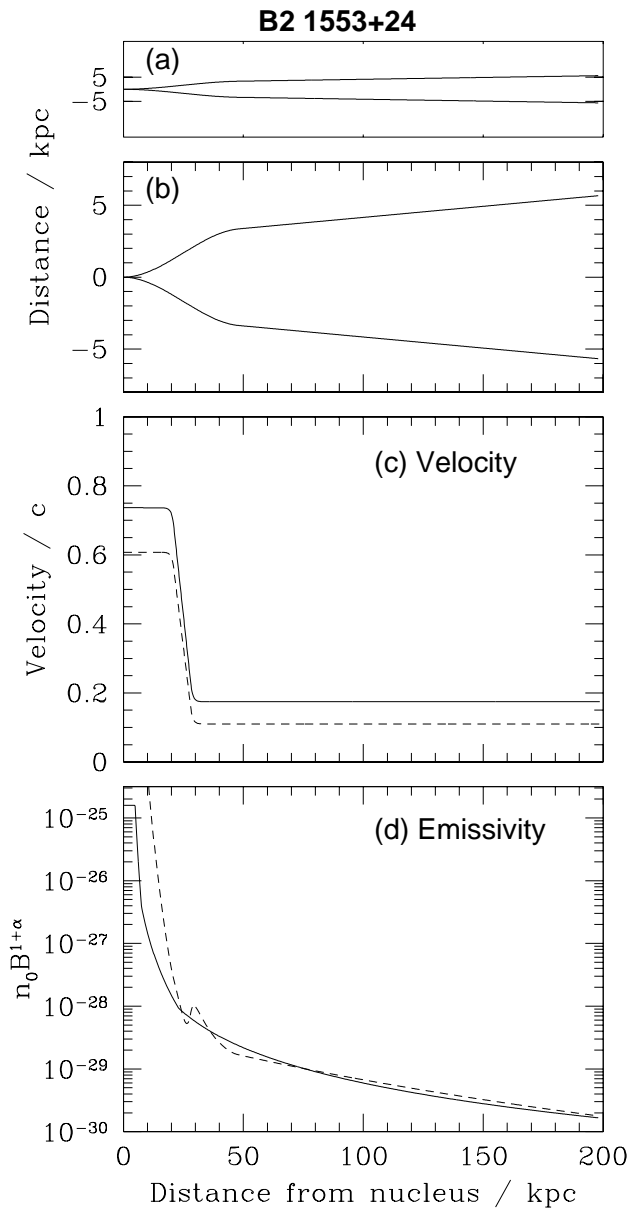}
\caption{Profiles of intrinsic parameters along the jets of B2\,1553+24 in
the jet frame. (a) and (b) the shape of the jet edge. Panel (a) shows the
jet with equal scales in both axes to give an idea of its real
shape. Panel (b) has the width scale expanded by a factor of 6 to
illustrate the variations in collimation. (c) the velocity profile along
the jet axis (solid line) and jet edge (dashed line). (d) the on-axis
profile of $n_0 B^{1+\alpha}$ derived from the emissivity, with $n_0$ in
m$^{-3}$ and $B$ in T. Solid line: model; dashed line, adiabatic
approximation with the magnetic-field structure expected from flux
freezing, normalized to match the free model at large distances.
\label{fig:1553profiles}}
\end{figure}

\subsubsection{Geometry and angle to the line of sight}
\label{1553angle}

Our best-fitting model has the jet axis at $7.7^\circ$ to the line of
sight with an error of $\pm1.3^\circ$. This implies that the jet is
has a very narrow `pencil' shape, with a diameter of only
$\approx$10\,kpc at 200\,kpc from the nucleus
(Fig.~\ref{fig:1553profiles}a). The jet is well collimated very close
to the nucleus, flares to a maximum opening angle of
$\approx$6$^\circ$ between 15 and 35 kpc from the nucleus and then
recollimates into a cone with an opening angle of
$\approx$0.75$^\circ$ (Fig~\ref{fig:1553profiles}b).

A near end-on orientation is qualitatively consistent with other
measures (Section~\ref{1553intro}) and the source was
indeed selected to be close to the line of sight. Nevertheless, the
extremely small value of $\theta$ implies the existence of a large,
missing population of side-on counterparts to B2\,1553+24 in the B2
sample. We address this topic in Section~\ref{1553incl}.

\subsubsection{Velocity field}
\label{1553velocity}

Fig. \ref{fig:1553profiles}(c) shows the model velocity profiles on-axis and
at the jet edge ($s=1$) for  B2\,1553+24. Within 20\,kpc of the nucleus,
the data are consistent with a constant value of $\beta = 0.74$, but the
counter-jet is invisible in our images and we can only set a limit of
$\beta \ga 0.7$. Between 20 and 30\,kpc from the nucleus, the jet
decelerates very rapidly and at large distances the velocity tends to an
asymptotic value of $\beta = 0.17\pm0.04$.

The variation in velocity transverse to the jet axis close to the nucleus
is only constrained by the lack of a visible counter-jet, which indicates 
that the velocity at the jet edge must be $\beta > 0.5$ ($>0.7$ times
the on-axis value). Further from the nucleus, our best-fitting model 
has a ratio of edge to on-axis velocity $\approx$0.6, although the error
is high. As in B2\,0326+29, the velocities are too low to produce
significant variations in Doppler factor across the jets unless the
edge/on-axis ratio $\la$0.2.

\subsubsection{Emissivity}
\label{1553emissivity}

The on-axis profile of $n_0 B^{1+\alpha}$ is shown by the solid line
in Fig. \ref{fig:1553profiles}(d). Within $\approx$5\,kpc of the
nucleus the emissivity is constant, the dramatic brightening of the
jet in this region being produced by the rapid expansion alone: no
discontinuity is required. Further from the nucleus, the profile falls
as a power-law, initially very steeply (index $\approx$8) but
flattening off as the distance increases. Between the start of the
emissivity decrease and the end of the modelled region the emissivity
drops by five orders of magnitude. In the regions of the jet where we
have adequate transverse resolution, the emissivity falls to
$\approx$0.25 of its on-axis value at the jet edge with an error of
roughly a factor of two.

\subsubsection{Magnetic-field structure}
\label{1553magnetic}

\begin{figure}
\includegraphics[width=8.2cm]{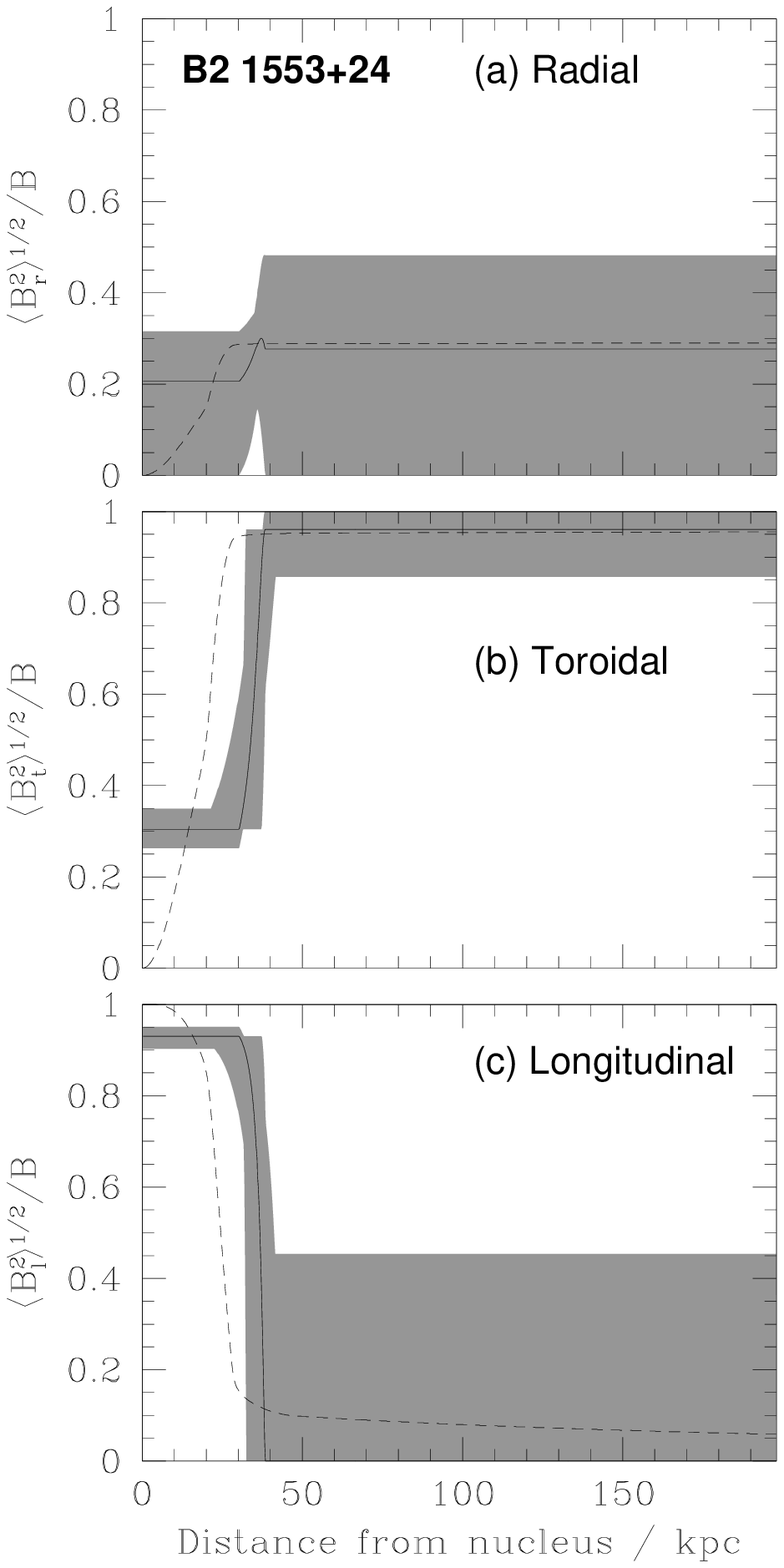}
\caption{Profiles of (a) radial, (b) toroidal and (c) longitudinal
components of the magnetic field along the streamlines in
B2\,1553+24. Solid line: best-fitting model; shaded area: error on the
model derived from the limits in Table \ref{Params}; dashed line: profile
expected on-axis, in the absence of shear if the magnetic field is
frozen into the plasma. The latter curves are normalized to match the model
values at 38 kpc from the nucleus.
\label{fig:1553bprofiles}}
\end{figure}

Profiles of the rms field components along the jets are shown by the solid
lines in Fig. \ref{fig:1553bprofiles}.  As in B2\,0326+39, there is no
variation with streamline index. The shaded areas show the variations
allowed by the error estimates in Table \ref{Params}. Within 30\,kpc of
the nucleus, the field is mostly longitudinal; between 30 and 40\,kpc
there is a rapid transition to a toroidally dominated structure which is
maintained over the rest of the modelled region (although the error
estimates do allow a significant proportion of longitudinal field to
remain).  The best fit for the fractional radial component is close to a
constant value of $\approx$0.2, but the data are also consistent with a
complete absence of radial field. A pure toroidal field would fit the
polarization data for $z > 40$\,kpc.

%-------------------------- ****** DISCUSSION ****** -----------------------------%
\section{Discussion}
\label{discuss}
\subsection{Adiabatic models}
\label{adiabatic}

The simplest physical picture of the evolution of the emissivity along
a jet assumes that the radiating particles only lose energy
adiabatically (synchrotron and inverse-Compton losses being negligible
by comparison), that there are no dissipative processes such as
particle acceleration or field-line reconnection, and that the
magnetic field is convected passively with the (laminar) flow. Analytical
equations for the emissivity in this case were first derived by
\citet{Burch79}; these were generalized by \citet{Baum97} to describe
a relativistic, decelerating flow with purely perpendicular or
parallel field.  We use the latter expressions, modified to calculate
the magnetic-field evolution self-consistently as in \citet{Laing04},
to estimate the emissivity profiles expected for our models of jet
shape and velocity.

We refer to the fits derived in the previous sections and by LB as {\em
free models} in order to distinguish them from {\em adiabatic models},
following the terminology of \citet{Laing04}.

\subsubsection{Magnetic-field profiles}
\label{freeze}

\citet{Baum97} showed that the variations of the
magnetic field components in the quasi-one-dimensional approximation are:
\begin{eqnarray*}
B_r &\propto& (x\beta\Gamma)^{-1} \\
B_t &\propto& (x\beta\Gamma)^{-1} \\
B_l &\propto& x^{-2}              \\
\end{eqnarray*}
in the absence of shear ($x$ is again the jet radius). Given the field
components at one point in the jet, we can then predict their
evolution from the profiles of radius and velocity given in the
previous section.

The predicted evolution of the field components in B2\,0326+39 and
B2\,1553+24 is shown by the dashed lines in Figs~\ref{fig:0326bprofiles}
and \ref{fig:1553bprofiles}, respectively. The initial conditions have
been chosen to match the free models at the distances from the nucleus
given in the figure captions (note that the field is forced to be
longitudinal at the nucleus).
 
The qualitative evolution of the field components is consistent with
the free models for both sources: a decrease in the longitudinal
component is accompanied by an increase in the toroidal component.
Rapid flaring and deceleration act in the same sense, leading to a
large decrease in the relative fraction of longitudinal field, as
observed. There are significant quantitative discrepancies, however,
particularly in the regions where the jets decelerate. In both
sources, the transition from longitudinal to transverse field is
predicted to be less abrupt than is seen in the free models.  This
discrepancy is qualitatively consistent with the effects of velocity
shear due to a transverse velocity gradient, which will act to
increase the longitudinal component and to delay the onset of the
longitudinal to transverse transition.  This cannot be the whole
story, however, as there are also large differences between the
variations of the toroidal and radial components. The former
increases, as expected, but the latter stays roughly constant: in the
quasi-one-dimensional approximation, their ratio should remain
constant. Similar problems occur in 3C\,31 (LB; \citealt{Laing04}).

\citet{KO89,KO90} showed that the variation of the rms
toroidal and longitudinal field components with distance could be very
different from that predicted by the simple adiabatic approximation if
the flow is turbulent, as is widely believed (e.g.\
\citealt{Bic84,DeY96}). Given that the gross departures from the
flux-freezing predictions occur where the jets are fast, or are
decelerating rapidly, it seems likely that turbulence affects both the
magnitude and the configuration of the field in these regions.  

\subsubsection{Emissivity profiles}
\label{ademiss}

The emissivity function $\epsilon$ can be written in terms of the magnetic
field $B$, both as defined in Section~\ref{Emissivity}, as:
\begin{eqnarray*}
\epsilon \propto (x^2\beta\Gamma)^{-(1+2\alpha/3)}B^{1+\alpha}
\end{eqnarray*}
\citep{Baum97,Laing04}.  $B$ can be expressed in terms of the
parallel-field fraction $f = \langle B_l^2 \rangle ^{1/2}/B$ and the
radius $\bar{x}$, velocity $\bar{\beta}$ and Lorentz factor
$\bar{\Gamma}$ at some starting location using equation 8 of \citet{Laing04}:
\begin{eqnarray*}
B \propto \left[ f^2\left(\frac{\bar{x}}{x}\right)^4
+ (1 - f^2)\left(\frac{\bar{\Gamma}\bar{\beta}\bar{x}}{\Gamma\beta
x}\right)^2\right ]^{1/2}
\end{eqnarray*}
We can therefore predict the emissivity using our fitted jet width and
velocity together with an estimate of the parallel-field fraction $f$.
The resulting emissivity profiles for B2\,0326+39 and B2\,1553+24 are
shown in Figs~\ref{fig:0326profiles}(c) and \ref{fig:1553profiles}(d),
respectively. The solid lines show the emissivity profiles from our free
model fits and the dashed lines the self-consistent adiabatic profiles,
normalized to match at large $z$.

In both objects the adiabatic models agree poorly with the free models
where the jet velocities are high. The emissivity falls off much too
rapidly in the adiabatic models, just as in 3C\,31 (LB). This is not a
surprise, for the following reasons:
\begin{enumerate}
\item Velocity shear (required by the free model) has been neglected.
\item The magnetic-field evolution is more complicated than expected from
simple flux-freezing in an axisymmetric laminar-flow model, even if shear
is included (Section~\ref{freeze}) and turbulence may dominate the
field evolution \citep{KO89,KO90}.
\item Where the jets are fast, we see complex, small-scale,
non-axisymmetric structures, indicating that the flow is not laminar. 
\item In B2\,1553+24, there is optical synchrotron emission from the base
of the main jet \citep{Parma03}, implying continuing particle
acceleration.
\end{enumerate}

Further from the nucleus, where the velocity has a low and roughly
constant value, the adiabatic profile matches the free model very well in
both objects. In B2\,0326+39 this is within a region 4 to 12\,kpc
(Fig.~\ref{fig:0326profiles}).  In B2\,1553+24, the region of low, constant
velocity extends from 30 to 200\,kpc and the adiabatic profile agrees very
well with the free model over approximately 1.5 orders of magnitude in
emissivity (Fig.~\ref{fig:1553profiles}).  This suggests that the outer
jets in this source are modelled surprisingly well as constant-velocity,
perpendicular-field, adiabatically-evolving flows.

\subsection{Sidedness profiles and reacceleration}
\label{acceleration}

\begin{figure*}
\includegraphics[width=17cm]{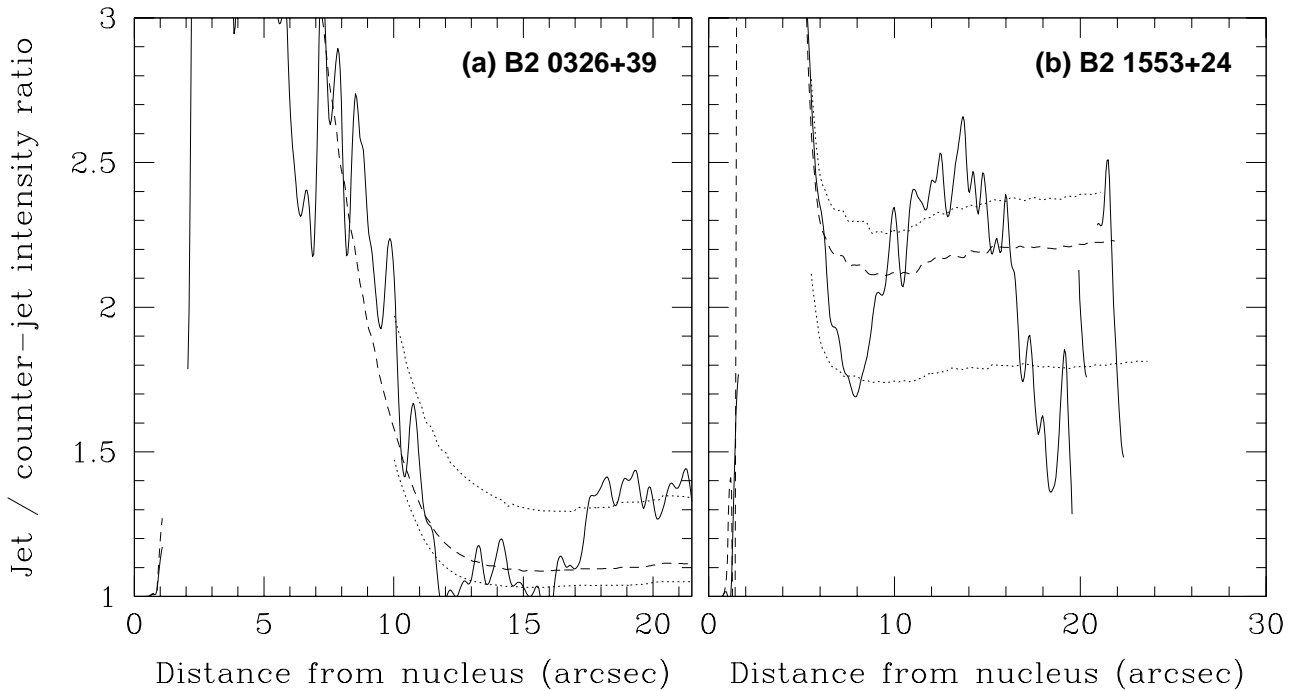}
\caption{Sidedness profiles of: (a) B2\,0326+39 at 0.5 arcsec resolution 
and (b) B2\,1553+24 at 0.75 arcsec resolution, averaging over all points at
the same distance from the nucleus. Solid line: data; dashed line: best
fit model; dotted lines: models with $\beta_0$ modified to fit the minimum
and maximum sidednesses in the lower-velocity region, as described in the text.
\label{avside}}
\end{figure*}

Our models, motivated by the gross features of the observed sidedness
ratios, assume monotonic deceleration. There are theoretical reasons
to expect jets to be reaccelerated by the pressure gradient of the
external medium at large distances from the nucleus provided that the
mass injection rate is not too large, for example if stellar mass loss
dominates the mass injection \citep{Komissarov94,Bowman96}.  The most
obvious effect of reacceleration is a small increase of sidedness
ratio with distance from the nucleus.  No such increase is obvious
from the ridge-line profiles or images of sidedness ratio
(Figs~\ref{fig:0326ls} and \ref{fig:1553ls}). We have therefore
averaged the sidedness ratios in rings of constant distance from the
nucleus in order to improve the signal-to-noise ratio
(Fig.~\ref{avside}).

This reveals considerable sidedness structure in the low-velocity
regions. Both sources show sidedness minima, at $\approx$14\,kpc from
the nucleus in B2\,0326+29 and at $\approx$8\,kpc in B2\,1553+24. The
minimum for B2\,0326+39 occurs roughly where \citet{Worrall00}
inferred that the synchrotron minimum pressure in the jets
\citep{Bridle91} becomes less than the pressure of the external
medium.

Fig.~\ref{avside} shows the average observed sidednesses as solid lines
and our best-fitting models as dashed lines. The dotted lines show the
profiles for models with the outer velocity parameters, $\beta_0$,
modified to fit the minimum and maximum sections of the observed profiles
in the outer region. The velocity ranges are $\beta_0 =$ 0.03 -- 0.25 for
B2\,0326+39 and 0.13 -- 0.19 for B2\,1553+24.  [Note that the model
sidedness profiles increase slightly with distance from the nucleus
because progressively larger areas of jet edge, which have lower
sidedness ratios than the centres, fall below the intensity blanking
threshold].

If the changes in sidedness result from acceleration, then they should
be associated with variations in the degree of polarization.  In
B2\,0326+39, the field structure can be roughly approximated by
two-dimensional field sheets with equal radial and toroidal components
(Section~\ref{0326magnetic}). The expected changes in polarization are
then straightforward to calculate \citep{Laing80}.  For an
acceleration from $\beta = 0.03$ to $\beta = 0.25$ between 16 and
18\,arcsec from the nucleus, as implied by a naive interpretation of
Fig.~\ref{avside}(a), $p$ should vary from 0.49 to 0.64 in the main
jet and from 0.45 to 0.38 in the counter-jet. These changes are not
observed (Figs~\ref{fig:0326lp}e and \ref{fig:0326avlp}). The velocity
increase is far larger than is expected from the effects of any
pressure gradient in the external medium and we also note that the fit
of an accelerating adiabatic model to the emissivity profile would be
significantly worse than that shown in Fig.~\ref{fig:0326profiles}.

In B2\,1553+24, it is more difficult to exclude acceleration as the cause
of the increase in sidedness ratio between 8 and 14\,arcsec from the
nucleus (Fig.~\ref{avside}b): the predicted variations in degree of
polarization are below the fluctuation level in Fig.~\ref{fig:1553lp}(e)
and we cannot average across the jet to reduce the noise; the velocity
increase required ($\beta_0 =$ 0.13 -- 0.19) is physically more reasonable
than in B2\,0326+39 and the adiabatic model fit to the emissivity is
comparable in quality to that in Fig.~\ref{fig:1553profiles}.

We conclude that there is no compelling evidence in favour of
reacceleration of the jets in either source, and that it is most unlikely
to be the sole cause of the increase in sidedness ratio with distance from
the nucleus in B2\,0326+39. Nevertheless, it is expected theoretically,
there are hints that it might occur, and it is consistent with the total
intensity and polarization data in B2\,1553+24.

\subsection{The appearance of B2\,1553+24 at other angles to the line
  of sight}
\label{1553incl}

We infer that the angle to the line of sight for B2\,1553+24 is
$\approx$8$^\circ$. Given that the B2 sample from which it is drawn is
selected at the low frequency of 408\,MHz, the usual assumption is
that the emission is isotropic and therefore that the distribution of
jet orientations for its members should be random. The source should
therefore have many ($\sim$100) counterparts of comparable total
luminosity and larger $\theta$, but the entire sample of B2 radio
galaxies with jets defined in Tables\,1 and 2 of \citet{Parma87} has
43 members. A further concern is that the counterparts must be very
large. The length of the main jet is at least 60\,arcsec
(Fig.~\ref{fig:1553.montage}a), corresponding to 388\,arcsec
(340\,kpc) at $\theta = 60^\circ$, the median angle to the line of
sight. This is comparable in size with the longest jet in the sample,
in NGC\,315 \citep{Willis81}, and far larger than the median
($\approx$30\,kpc; \citealt{Parma87}). There is cause for suspicion
unless: (a) the side-on counterparts of B2\,1553+24 are not members of
the B2 sample or (b) sources in that sample have linear sizes far
larger than previously realised.

\begin{figure*}
\includegraphics[height=17cm,angle=-90]{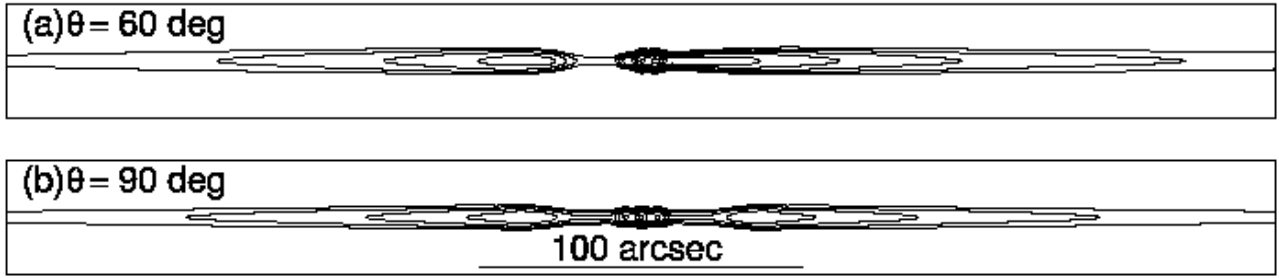}
\caption{The brightness distribution of the model for the jets in
  B2\,1553+24 at angles to the line of sight of: (a) $\theta =
  60^\circ$ and (b) $\theta = 90^\circ$. The plots cover 394\,arcsec
  (346\,kpc) in projection on the sky. This corresponds to a distance
  from the nucleus along the jet axis of 200\,kpc in both directions
  for $\theta = 60^\circ$, i.e.\ to the size of the modelled
  region. The contour levels are 1, 2, 4, 8, 16, 32, 64, 128, 256, 512
  $\times$ 20$\mu$Jy/beam and the model images have been convolved
  with a 2.75-arcsec Gaussian beam.\label{fig:1553los}}
\end{figure*}

In order to understand potential selection effects, we have computed
the appearance of the model brightness distribution for B2\,1553+24 at
various angles to the line of sight. We show the results for the
median angle to the line of sight for an isotropic sample ($\theta =
60^\circ$) and for a source in the plane of the sky ($\theta =
90^\circ$) in Fig.~\ref{fig:1553los}. The area chosen for the plots
corresponds to the modelled region for the $60^\circ$ case.  As well
as being very long, the jets appear narrow, straight and (except for a
small region around the nucleus) extremely faint.  The total flux
density from the inner 200\,kpc of the model jets (measured along the
jets and excluding the core) is constrained to be 44\,mJy for the
$\theta = 7.7^\circ$ model but is 27\,mJy for $\theta = 60^\circ$.
There are three important selection effects:
\begin{enumerate}
\item The extended flux of the jets shows significant Doppler boosting
  at small $\theta$. Given that there is little lobe emission in
  B2\,1553+24 \citep{deRuiter93,NVSS}, the jets probably dominate the flux at
  low radio frequencies, so end-on sources of this type will be
  selected preferentially in the B2 survey.  The integral source count
  $N(>S) \propto S^{-3/2}$ for the luminosities and redshifts in
  question, so the size of the parent sample is effectively increased
  by a factor of $(44/27)^{3/2} \approx 2$ if isotropic lobe emission is ignored.
\item Except for a small region around the nucleus, the model jets are
  faint. Further than 50\,arcsec from the nucleus, the surface
  brightness is $\la$0.16\,mJy/beam area at 8.4\,GHz with a beam of
  2.75\,arcsec FWHM. This scales to 0.47\,mJy/beam area at 1.4\,GHz
  for a spectral index $\alpha = 0.6$, comparable with the
  detection threshold for typical VLA images of sources in the sample
  presented by \citet{Fanti86,Fanti87}. In any case, emission on scales
  $\ga$120\,arcsec would not have been imaged reliably in existing VLA
  observations of the B2 sample (C configuration at 1.4\,GHz;
  \citealt{Fanti87}).  The combination of these two effects makes it
  extremely unlikely that the outer jets would have been detected.
\item Sources of large angular size with flux densities close to the
  limit of the catalogue could have been missed by the original B2
  survey, which measured peak rather than integrated flux
  density. We have estimated the magnitude of this effect by convolving
the model brightness distribution for $\theta = 60^\circ$ with the beam of
the B2 survey (3 $\times$ 10\,arcmin$^2$). If the short axis of the beam
is parallel to the jet axis, the ratio of peak/total flux is 0.73; if the
long axis of the beam is aligned, the ratio becomes 0.95. 
A few sources could therefore have been missed (primarily those orientated E-W).
\end{enumerate}
We conclude that the side-on counterparts of B2\,1553+24 could have
escaped identification, either because they were missed in the original
survey or because their angular sizes have been greatly
underestimated. Morganti \& Parma (private communication) and Ledlow
\& Owen (2004; in preparation) have made more sensitive observations
of radio galaxies in the B2 sample using the WSRT and the VLA in D
configuration and have shown that a significant fraction of them have much
longer radio jets than have previously been reported, extending
many 100's of kpc or even further. We suggest that these form part of
the missing population.

\subsection{Similarities and differences between B2\,0326+39, B2\,1553+24 and 3C\,31}
\label{compare}

In later papers, we will present a re-analysis of 3C\,31 along with models
of two further objects, all with the same functional descriptions of
velocity, emissivity and magnetic field. This will allow a full comparison
between them; here we make a few general observations about the
similarities and differences between B2\,0326+39, B2\,1553+24 and 3C\,31.

Notable similarities are: 
\begin{enumerate}
\item All three objects can be modelled successfully by a decelerating jet model.
\item The jets are all initially well collimated, then flare before
recollimating and expanding conically.
\item Close to the nucleus the jet velocity on-axis is consistent with a value of 
$\beta \approx 0.8$ in all cases.
\item Deceleration occurs over a region of $\approx$10\,kpc to an
outer velocity of $0.0 \la \beta \la 0.2$.
\item The ratio of edge to on-axis velocity is consistent with a constant
value of $\approx$0.7 everywhere, although in some places this is poorly
constrained.
\item The exponent of the emissivity index is similar in the outermost
modelled regions of all the jets.
\item At the jet edge the emissivity falls to $\approx$0.25 of its on-axis
value.
\item The longitudinal field component dominates in the high-velocity regions
close to the nucleus and the toroidal component in the outer parts,
qualitatively but not quantitatively as expected from flux freezing.
\item The variation of the radial field component is more complex than
predicted by simple flux-freezing models. This effect cannot simply be
due to velocity shear in a laminar, axisymmetric flow.
\item Close to the nucleus (in the high-velocity and deceleration
regions), the variation of emissivity with distance from
the nucleus is far less rapid than predicted by the adiabatic
approximation.
\item In the low-velocity outer regions, the emissivity variation is
reasonably well fitted by a quasi-one-dimensional adiabatic model.
\end{enumerate}

The three objects were expected to have a wide range of angles to the line
of sight; this is confirmed by our modelling. In addition, there are some
clear differences:
\begin{enumerate}
\item The conical outer region of 3C\,31 has a large opening angle and
is centred on the nucleus.  The outer regions of the other two objects
are closer to cylindrical, with very small opening angles and vertices
far behind the nucleus. [Note that 3C\,31 does recollimate 
outside the modelled region (LB)].
\item The velocity in the outer region of 3C\,31 was modelled by LB as
decreasing monotonically with distance from the nucleus (as required by the
variation of sidedness ratio). In the B2 sources, it has an approximately
constant asymptotic value.
\item The close connection between the jet geometry and the forms of the velocity
and emissivity profiles inferred for 3C\,31 is not general.
\item The velocity in B2\,1553+24 remains at $\beta \approx 0.8$ until
$\approx$20\,kpc from the nucleus (cf.\ $\approx$2\,kpc in the
other two objects).
\item There is no need for a discontinuity in emissivity in
B2\,1553+24: the brightening of the jet is consistent with expansion
at constant emissivity.
\item Although the biggest single field component at large distances is
always toroidal, there are significant differences in the 
details of the field-component evolution: 
3C\,31 has a mixture of longitudinal and toroidal components; B2\,0326+39
has toroidal and radial components and B2\,1553+24 has almost pure
toroidal field.
\item In 3C\,31, the radial field component in the flaring region
increases towards the edge of the jet. There is no evidence for this
effect in the B2 sources, although it cannot be excluded in 
B2\,0326+39, where the signal-to-noise ratio is low.
\item The adiabatic approximation describes the emissivity evolution in
B2,1553+24 over a much larger fraction of the jets than in the other two
sources. This may be because it applies most accurately in the low-velocity
outer region, which is relatively longer in this source. 
\end{enumerate}

%------------------------- ****** FURTHER WORK ****** ------------------------%
\section{Summary and further work}
\label{ssfw}

\subsection{Summary}
\label{summary}

\subsubsection{General}

We have shown that the total and linearly-polarized synchrotron emission
from the jets of the FR\,I radio galaxies B2\,0326+39 and B2\,1553+24 can
be modelled successfully on the assumption that they are axisymmetric,
intrinsically symmetrical, relativistic, decelerating flows.  The models
are based on those developed by LB for 3C\,31, with modifications to the
functional forms assumed for geometry, velocity, emissivity and magnetic
field. In particular, we found that the variations of intrinsic parameters
were not directly coupled to the geometry of the jets as defined by their
outer isophotes.

\subsubsection{Geometry}

In the {\em flaring region} close to the nucleus, the jet radius $x$ can
be fitted by a polynomial of the form $x = a_2 z^2 + a_3 z^3$, where $z$
is the distance from the nucleus. In the {\em outer region} at larger
distances, the jets are conical, with opening angles sufficiently small
that they are almost cylindrical. The inferred angles to the line of
sight are $64^\circ \pm 5^\circ$ for B2\,0326+39 and $7.7^\circ \pm
1.3^\circ$ for B2\,1553+24.

\subsubsection{Velocity}

Close to the nucleus the jets have high and approximately constant
velocities $\beta \approx 0.8$.  They decelerate rapidly over distances of
$<10$\,kpc to a constant asymptotic outer velocities $\beta \la 0.2$. The
velocities at the jet edges are consistent with $\approx$0.7 of their
on-axis values everywhere.

\subsubsection{Emissivity}

The emissivity profiles along the jets can be divided into discrete
regions, each with a power-law variation. The power-laws tend to be
flatter further from the nucleus, although regions close to the nucleus
(where knots of emission are seen) are modelled as bright plateaux with
rapid decreases in emissivity at their ends. In the low-velocity
regions after the jets have decelerated, the emissivity profiles are
consistent with a simple, quasi-one-dimensional adiabatic model; closer to
the nucleus, the adiabatic approximation predicts too steep a decline of
emissivity with distance. The fractional emissivity at the edge of the
jets is $\approx$0.25 everywhere.

\subsubsection{Magnetic Field}

Although the observed polarization structures of the two objects appear
radically different, the main features of the magnetic-field evolution are
similar in both cases: the principal differences result from projection.
In both objects, the field is dominated by the longitudinal component in
the region close to the nucleus; further out, the largest single component
is toroidal. The radial field components are $\approx$0.4 and $\approx$0.2
of the total in B2\,0326+39 and B2\,1553+24, respectively.  No evidence
for transverse structure in the field-component ratios is seen in either
source. Evolution from longitudinal to transverse field is qualitatively
consistent with flux-freezing in an expanding, decelerating jet, but there
are complications.  First, the ratio of radial to toroidal field (which
should remain roughly constant even if there is velocity shear) varies
significantly along the jets. Second, the details of the component
evolution are quantitatively inconsistent with the flux-freezing picture.

\subsection{Further work}
\label{further work}

We have modelled 3C\,31 with the modified functional descriptions used
here in order to allow a quantitative comparison with other objects.
As expected from the variation of sidedness ratio with distance from
the nucleus (LB), the jet velocity in 3C\,31 does not reach an
asymptotic value in the modelled region. We have shown, however, that
use of the new velocity law allows as good a fit to the observations
as that achieved by LB and that their conclusions are not
significantly modified.  We have also successfully applied the model
to the well-known radio galaxies NGC\,315 \citep{Venturi93} and
3C\,296 \citep{Hardcastle97}. These results will be presented in
future papers.

When suitable X-ray observations are available, we will apply the
conservation-law approach developed by \citet{LB2} to infer the energy and
momentum fluxes of the modelled jets and their variations of pressure,
density and entrainment rate with distance from the nucleus. Our eventual
aim is to replace the empirical descriptions of internal quantities with
more physically derived ones. Initially, we will apply the more
sophisticated adiabatic model developed by \citet{Laing04} to the outer
regions of the jets and in particular to B2\,1553+24, where the simple analysis
of Section \ref{adiabatic} suggests that the adiabatic approximation holds
over a large range of distances. We then plan to 
incorporate the effects of synchrotron and inverse Compton losses in a
self-consistent way, to model the observed emission from radio to X-ray
wavelengths and to quantify the particle-acceleration processes in FR\,I
radio jets.

%--------------------------- ****** THE END ****** ------------------------------%
\section*{Acknowledgments}

JRC acknowledges a research studentship from the UK Particle Physics
and Astronomy Research Council (PPARC). The National Radio Astronomy
Observatory is a facility of the National Science Foundation operated
under cooperative agreement by Associated Universities, Inc. We thank
Paola Parma, Raffaella Morganti, Frazer Owen and Michael Ledlow for
communicating results in advance of publication, Alan Bridle for
comments on the text and the referee for constructive suggestions.

\label{lastpage}
\end{document}